\documentclass[preprint,eqsecnum,floats,aps]{revtex4}
\usepackage{graphicx}
\usepackage{bm}
\begin{document}
\include{def}
\def\v#1{\mbox{\boldmath$#1$}}
\newcommand{\smod}{******START MODIFICATION****** : }
\def\ket#1{|#1 \rangle}
\def\bra#1{\langle #1|}

\vspace{2cm}
\title{Nonmesonic weak decay of $\Lambda$ hypernuclei within a nuclear matter framework.}
\author{E. Bauer}

\affiliation{
Departamento de F\'{\i}sica, Universidad Nacional de
La Plata,\\
C. C. 67, 1900 La Plata, Argentina}
\affiliation{
Instituto de F\'{\i}sica La Plata,
CONICET, 1900 La Plata, Argentina}

\begin{abstract}
The nonmesonic weak decay of $\Lambda$ hypernuclei is studied
using nuclear matter. We have developed a formalism which gives a
microscopic interpretation of the process of emission of particles
originated in this decay. More specifically, our scheme provides a
unified treatment of $\Gamma(\Lambda N \rightarrow NN$) and
$N_{N}$ ($N_{NN}$), the $\Lambda$ non-mesonic weak decay widths
and the number of emitted particles (pair of particles) of kind
$N$ ($NN$), respectively. We have also evaluated for the first
time the quantum interference terms between the $n$- and
$p$-induced weak decay amplitudes. Explicit expressions for
$N_{N}$ and $N_{NN}$ are shown within the ring approximation.
Using the local density approximation together with the ring
approximation, we report results for the decay of the
$^{12}_{\Lambda}C$ hypernucleus. We have obtained values for the
ratio $N_{n}/N_{p}$ ($N_{nn}/N_{np}$) in the range $1.4-1.6$
($0.2-0.3$), where no energy threshold have been employed. The
$n$- and $p$-induced interference terms modify in less than $\sim
3 \, \%$ these results.

\vspace{2.6cm}


\vspace{.5cm}

\end{abstract}

\maketitle

\newpage

\section{INTRODUCTION}
\label{INTR}
The physics of hypernuclei has started in 1952 by the observation
of the first hypernucleus through its decays \cite{da53}.
Since then, the subject has grown and in the present
contribution we are concern with a particular topic of the
hypernuclei physics: the weak decay of $\Lambda$-hypernuclei. For
review articles one can see \cite{ra98,al02}. Many
theoretical and experimental~\cite{mo74,ha01,sa05,ki02,ok04,bhang,outaVa,kang06,kim06},
efforts have been developed to understand
the physics beyond this decay. The two main decay mechanisms are:
the Mesonic decay ($M$), where the decay reaction is
$\Lambda \rightarrow \pi N$, that is, the $\Lambda$ inside the
hypernucleus decays into a pion and a nucleon. The second is named
Non-Mesonic decay ($NM$). There are several reactions which contribute to
the $NM$-decay and the simplest one is $\Lambda N \rightarrow N N$.
Any of both reactions can occur when a hypernucleus decays. However,
the kinematical conditions of the emitted nucleons are very
different between the mesonic and the non-mesonic decays.
While in the mesonic decay, the momentum carried by
the nucleon is of the order of 100~MeV/c, in the $NM$ case, this
value is 400~MeV/c (assuming that the two emitted nucleons have the same
momentum). This has a strong effect on the values of the
corresponding decay widths: the mesonic one ($\Gamma_{M}$), is
inhibited by the Pauli blocking. In this work, we
focus on the non-mesonic decay width ($\Gamma_{NM}$) of
$\Lambda$-hypernuclei.

For the $NM$-decay the transition rates can be
stimulated either by protons, $\Gamma_p \equiv
\Gamma(\Lambda p \rightarrow n p)$, or by neutrons,
$\Gamma_n \equiv \Gamma(\Lambda n \rightarrow n n)$.
The total $NM$-decay rate is then $\Gamma_{NM} =
\Gamma_n + \Gamma_p $. The theory fairly accounts
for the experimental values of the total transition rate.
There are two quantities which remain not fully understood yet.
The first one is the ratio $\Gamma_{n/p} \equiv \Gamma_n /
\Gamma_p$, which is discussed soon. The other one is
the asymmetry of the protons emitted in the $NM$-decay
of polarized hypernuclei. In this case, the data indicates
a value closer to zero, while the theory predicts a large negative number.
In the present work, however, we will not deal with the asymmetry.

The ratio $\Gamma_{n/p}$ is evaluated using many theoretical models.
The first microscopic scheme has been proposed by
Adams \cite{ad67}, who has used the  nuclear matter framework, one pion
exchange model (OPE), the $\Delta T = 1/2$ piece of the $\Lambda N
\pi$ coupling and short range correlations (SRC). The
OPE produces ratios in the interval 0.05-0.20,
well below data. At variance, $\Gamma_{NM}$ is well
reproduced by OPE. From this point, a huge theoretical
effort has been devoted to find a formalism which increases
$\Gamma_{n}$ and decreases $\Gamma_{p}$.
Let us group them as the ones which care about: $i)$
the transition potential: the OPE can be improved by considering
heavier mesons than the pion.
This $\Lambda N \rightarrow N N$--transition potential is
known as one meson exchange model (OME) and it has been employed in
several works \cite{mc84,du96,pa97,ji01,pa02,it02}.
$ii)$ The inclusion of interaction terms that violates the isospin $\Delta T = 1/2$ rule
has been considered in \cite{al02,mal94,ino94,go97,pa98,sa00}.
$iii)$ the addition of two-body induced decay
mechanism stemming from ground state correlation of the
hypernuclei \cite{al91,ra94,ra97,al00b,al00,ba04}.
In point $ii)$, it should be observed that
the quark degree of freedom allows for  both
$\Delta T = 1/2$ and $\Delta T= 3/2$ transitions.
All theoretical models consistently obtain values
for $\Gamma_{n/p}$ smaller than 0.5 for medium and
heavy hypernuclei. In the experimental side, all measurements
suggest values greater than 0.5. This discrepancy is usually called the
$\Gamma_{n/p}$-puzzle. Recently, it has been suggested that the
origin of this inconsistency comes from the ambiguities
in the interpretation of data, rather than in
our poor understanding of the weak decay
mechanisms~\cite{ra97,ga03,ga04,ba06}. In these works, the intranuclear
cascade code (INC) has been developed, which
is a semi-phenomenological approach where first the $\Lambda$-weak
decay is evaluated microscopically and afterwards the nucleons
produced in the decay are followed in a semi-classical manner until
they leave the nucleus. By means of this emulation of the physical
conditions of the hypernuclear decay, a more accurate agreement between
the theoretical results and the data is achieved.

Beyond all these theoretical and experimental labor, the effect
of the nuclear structure on the calculation of $\Gamma_{NM}$ is
less or even poorly discussed.
Oset and Salcedo \cite{os85} have developed the polarization
propagator method (PPM), which allows a unified
treatment of the mesonic and the nonmesonic-decay channels.
Nuclear correlations are included through the ring approximation,
which is the direct part of the RPA. The PPM has been
further developed in Refs. \cite{ji01,ra94,ra97,al00b}.
In a similar spirit of the PPM, Alberico et al.~\cite{al00}
have employed the bosonic loop
expansion (BLE) formalism, which is  particularly suitable
when more than two nucleons emerge from the
disintegration process. In practice and until now, the
strong potential has entered into the evaluation of $\Gamma_{n}$ and
$\Gamma_{p}$, only by means of the ring approximation and
restricted to a $(\pi+\rho)$-potential with the addition of the
Landau-Migdal $g'$-parameter.

In the present contribution, we address the problem of the theoretical
interpretation of data. This is done within a fully microscopic model,
which, as a first step, is restricted to the ring approximation.
We pay special attention to the roll of the quantum interference
terms between the $n$- and $p$-induced weak decay amplitudes, which
are evaluated in this work for the first time.
The paper is organized as follows. In Section II, we discuss the
way in which $\Gamma_n/\Gamma_p$ is extracted from the experimental
measurements. In Section III, we propose a
model for the physical observables $N_{N}$ and $N_{NN}$.
In Section IV, explicit expressions
for these observables are given within the ring approximation. Finally,
in Sections V and VI, some results and conclusions are given.

\newpage

\section{THE EXTRACTION OF THE RATIO $\Gamma_n/\Gamma_p \,$ FROM DATA}
\label{puzzle}

Some recent works suggest a solution of the $\Gamma_n/\Gamma_p \,$-puzzle (see
\cite{ga03,ga04,ba06}). However, in this section we do not want to
go through this explanation, but to call the attention on some aspects
about the extraction of $\Gamma_n/\Gamma_p$ from data.
Therefore, let us start by presenting a simplified version
of the way in which the so-called experimental value of $\Gamma_n/\Gamma_p$
is obtained. This brief outline does not pretend to
be complete. For a more rigorous analysis, we refer the reader to the
experimental works \cite{mo74,ha01,sa05,ki02,ok04,bhang,outaVa,kang06,kim06}
and also to the above mentioned studies.
There are two equivalent definitions for the experimental
value of $\Gamma_n/\Gamma_p$ \cite{ga03,ga04},
\begin{eqnarray}
\label{gngpd1}
\frac{\Gamma_n}{\Gamma_p} & \equiv & \frac{1}{2} \, (
\frac{N^{\rm wd}_n}{N^{\rm wd}_p} - 1) \\
\label{gngpd2}
\frac{\Gamma_n}{\Gamma_p} & \equiv &
\frac{N^{\rm wd}_{nn}}{N^{\rm wd}_{np}}
\end{eqnarray}
where $N^{\rm wd}_{N}$ is the number of nucleons of kind $N$ produced
in the weak decay of the $\Lambda$. Analogously, $N^{\rm wd}_{N N}$
is the number of $N N$ pairs. Unfortunately, these primary particles
can not be measured. Once these particles are produced, in their way out
of the nucleus and due to collisions with other nucleons, they can change
energy, direction and charge. These interactions with
other nucleons are called Final State Interactions (FSI).
Perhaps, one of the simplest nuclear models to deal with the
FSI is the ring approximation, where the nuclear residual interaction
is replaced by a modified interaction built up from an infinite
sum of one particle-one hole bubbles.
The first order contributions to
the ring series for the $\Lambda$-decay are shown in Fig.~1,
where we have also drawn a diagram for $\Gamma_{N}$.
\begin{figure}
\begin{center}
    \includegraphics[width = .5\textwidth]{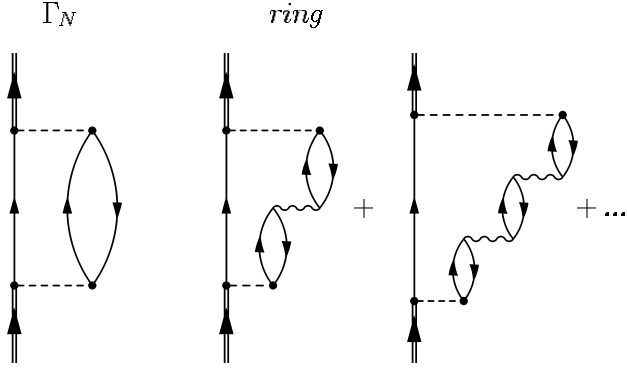}
\vskip 2mm
\caption{Goldstone diagrams for $\Gamma_{N}$ and
the ring series.
An up (down) going arrow represents a particle (hole), while
an arrow with a double line represents the $\Lambda$.
The dashed lines represent
the full $V^{\Lambda N}$-transition
potential and the
intermediate wavy lines represent the strong nuclear potential,
$V^{NN}$.}
\label{fig:fig1}
\end{center}
\end{figure}
As an example of the action of the FSI we discuss now the last diagram
in Fig.~1, which is of second order in the nuclear residual interaction.
To develop this example, let us
additionally consider that the primary decay is $\Lambda n  \rightarrow nn$. The
intermediate bubble can be either a $n n^{-1}$ or a $p p^{-1}$-pair.
We also suppose that this bubble is constituted by a $p p^{-1}$-pair and that it is
on the mass shell. The interpretation of this
contribution looks somehow confusing: the primary decay is
$\Lambda n  \rightarrow nn$, while the final state is $np$.
This $np$-pair is originated from the action of the strong interaction, after
the $\Lambda$-weak decay takes place. Then, and as it is already known, the emitted
particles depend on the FSI. The ring approximation is only one kind
among a huge set of FSI.
To express in general terms the effect of FSI,
in~\cite{ga03} it is written down a set
of equations which represents the number of nucleons of the kind $N$
(where $N$=$n$ or $p$),
outgoing the nucleus,
\begin{eqnarray}
\label{nn1}
N_{n} & = & N^{\rm 1Bn}_{n} \bar{\Gamma}_{n} + N^{\rm 1Bp}_{n} \bar{\Gamma}_{p} \\
\label{np1}
N_{p} & = & N^{\rm 1Bn}_{p} \bar{\Gamma}_{n} + N^{\rm 1Bp}_{p} \bar{\Gamma}_{p},
\end{eqnarray}
where $\bar{\Gamma}_{N} \equiv \Gamma_{N}/(\Gamma_{n}+\Gamma_{p})$. Here
$N_{n}$ ($N_{p}$) is the number of neutrons (protons) emerging from the
nucleus due to the $\Lambda$-weak decay, while
$N^{\rm 1B\textit{i}}_{n}$ ($N^{\rm 1B\textit{i}}_{p}$)
is the number of neutrons (protons) coming out of the nucleus which are
originated from the $\Lambda i \rightarrow n i$ weak decay ($i=n, \, p$).
In a similar way, for the $NN$-pairs,
\begin{eqnarray}
\label{nnn1}
N_{nn} & = & N^{\rm 1Bn}_{nn} \bar{\Gamma}_{n} + N^{\rm 1Bp}_{nn} \bar{\Gamma}_{p} \\
\label{nnp1}
N_{np} & = & N^{\rm 1Bn}_{np} \bar{\Gamma}_{n} + N^{\rm 1Bp}_{np} \bar{\Gamma}_{p} \\
\label{npp1}
N_{pp} & = & N^{\rm 1Bn}_{pp} \bar{\Gamma}_{n} + N^{\rm 1Bp}_{pp} \bar{\Gamma}_{p}.
\end{eqnarray}
where the interpretation of the new factors is self-evident.
In the last equation, the action of the FSI allows the emission of a $pp$-pair.
For simplicity, in this set of equations we have
limited ourselves to nucleons originated from the one-body ($1B$),
decay mechanism ($i.e.$ $\Lambda N \rightarrow NN$). The
two-body decay mechanism ($\Lambda NN \rightarrow NNN$) is important
in the analysis of data, but it is not required for the present discussion.

The Eqs.~(\ref{nn1}-\ref{np1}) and Eqs.~(\ref{nnn1}-\ref{npp1}),
are one of several proposals to interpret the experimental results.
Within this scheme, the
dependence on the weak vertex is embodied in $\Gamma_{n}$ and $\Gamma_{p}$.
Thus, the $N^{\rm 1B \, n (p)}_{N}$ and
$N^{\rm 1B \, n (p)}_{NN}$-factors depend
only on the strong interaction part of the
problem and the nuclear model employed to describe the nucleus.
Therefore, one working hypothesis of the INC calculation is
that the number of emitted particles can be drawn as a sum of
product functions: in these products, one factor ($\Gamma_{N}$)
contains the entire dependence on the weak vertex.
When the strong interaction is turned off, these factors reduce
themselves to the number of particles (pairs of particles)
of kind $N$ ($NN$) in the primary decay. In Table~I, we
quote these values.

\begin{table}[h]
\begin{center}
\caption{Multiplicative factors. The index $wd$, indicates
that the FSI are neglected. As in the one-body weak
decay of the $\Lambda$ there is no $pp$-pair, we have
$(N^{\rm 1Bn}_{pp})^{wd}$=$(N^{\rm 1Bp}_{pp})^{wd}$=0.}
\vspace{1cm}
\begin{tabular}{cccc}   \hline\hline
~~~$(N^{\rm 1Bn}_{n})^{wd}$~~~ & ~~~$(N^{\rm 1Bp}_{n})^{wd}$~~~ &
~~~$(N^{\rm 1Bn}_{p})^{wd}$~~~ & ~~~$(N^{\rm 1Bp}_{p})^{wd}$~~~ \\ \hline
2 & 1 & 0 & 1 \\ \hline\hline
~~~$(N^{\rm 1Bn}_{nn})^{wd}$~~~ & ~~~$(N^{\rm 1Bp}_{nn})^{wd}$~~~ &
~~~$(N^{\rm 1Bn}_{np})^{wd}$~~~ & ~~~$(N^{\rm 1Bp}_{np})^{wd}$~~~ \\ \hline
1  &  0  &  0  &  1  \\ \hline\hline
\end{tabular}
\end{center}
\end{table}
If we replace now
the numbers in Table~I, into Eqs.~(\ref{nn1}-\ref{nnp1}), we have,
\begin{eqnarray}
\label{nn10}
N^{\rm wd}_{n} & = & 2 \bar{\Gamma}_{n} +  \bar{\Gamma}_{p} \\
\label{np10}
N^{\rm wd}_{p} & = & \bar{\Gamma}_{p},
\end{eqnarray}
and
\begin{eqnarray}
\label{nnn10}
N^{\rm wd}_{nn} & = & \bar{\Gamma}_{n} \\
\label{nnp10}
N^{\rm wd}_{np} & = & \bar{\Gamma}_{p}
\end{eqnarray}
which lead to the Eqs.~(\ref{gngpd1}) and (\ref{gngpd2}). When FSI are present,
we obtain from Eqs.~(\ref{nn1}-\ref{nnp1}),
\begin{equation}
\label{gngpn}
\frac{\Gamma_n}{\Gamma_p}=
\frac{\displaystyle N^{\rm 1Bp}_n
-N^{\rm 1Bp}_p \frac{N_n}{N_p}}
{\displaystyle N^{\rm 1Bn}_p \frac{N_n}{N_p} -N^{\rm 1Bn}_n} ,
\end{equation}
and
\begin{equation}
\label{gngpnn}
\frac{\Gamma_n}{\Gamma_p}=
\frac{\displaystyle N^{\rm 1Bp}_{nn}
-N^{\rm 1Bp}_{np} \frac{N_{nn}}{N_{np}}}
{\displaystyle N^{\rm 1Bn}_{np} \frac{N_{nn}}{N_{np}} -N^{\rm 1Bn}_{nn}}.
\end{equation}

Basically, within the INC calculation the so-called experimental value for
$\Gamma_n / \Gamma_p$ comes from one of these expressions. The
Eq.~(\ref{gngpn}) is more frequently used than the
Eq.~(\ref{gngpnn}). However, in \cite{al02} (see also
\cite{ga04}), it is claimed that the last expression is more
convenient, because it reduces the quantum mechanical
interferences between $n$ and $p$-stimulated weak decays. From the
experimental point of view, the quantities which can be measured
are $N_n$, $N_p$, $N_{nn}$, $N_{np}$ and $N_{pp}$. Therefore,
within the INC calculation the ratio
$\Gamma_n / \Gamma_p$ is obtained indirectly.

In recent years, there have been many improvements in the experimental
side: new and more accurate measurements, independent counts of
neutron and protons are performed, etc.
The theoretical works have extensively analyzed the weak decay,
as stated in the Introduction. Within the INC calculation point of view,
the link between the quoted theory and the data,
are the $N^{\rm 1Bn (p)}_N$ (or $N^{\rm 1Bn (p)}_{NN}$)-factors.
The INC calculation is one of several proposals in the analysis of the
experimental points. These other models are found in the
experimental works themselves. The simplest approach is
the direct use of Eq.~(\ref{gngpd1}) or (\ref{gngpd2}),
a model which certainly oversimplified the issue.
Among the different models, the INC calculation is perhaps the
more elaborated one and its result suggests a solution for the
$\Gamma_n / \Gamma_p$-puzzle. For $^{12}_{\Lambda}C$
there is some discrepancy left, but the use of the INC calculation
makes the difference smaller than with
the employment of other schemes. Moreover, within the INC calculation
this difference lays within the uncertainty of the experiment.

Despite assuming that the results from the INC calculation are appropriate enough,
nevertheless, a fair question from the theoretical point of view
would be, what the microscopic interpretation of
the $N_N$ and/or $N_{NN}$-factors is. In addition, the
results from the INC calculation are reliable as far as the quantum interference
terms between the $n$- and $p$-induced weak decay widths are small.
From now on, we will propose an answer for these points,
which are the main subjects of the present contribution.
A microscopic model for $N_{N}$ and $N_{NN}$  is given in
the next Section. Additionally, in the same Section we
attempt to interpret the factors $N^{\rm 1Bn (p)}_N$ and $N^{\rm 1Bn (p)}_{NN}$,
from the INC calculation.
In Section IV, explicit expressions for all these factors
are given using the ring approximation.

\newpage
\section{GENERAL EXPRESSIONS FOR THE $N_{N}$ AND $N_{NN}$ FACTORS}
\label{NNNs}

The starting point to build up microscopic expressions for
$N_{N}$ and $N_{NN}$ are the Eqs.~(\ref{nn10}-\ref{nnp10}).
These are the values for $N_{N}$ and $N_{NN}$ when no FSI are
present. Therefore we add to these equations the action of the
FSI. To this end, we introduce the quantity $\Gamma_{i, i' \rightarrow j}$.
In this function, the
indices $i, \, i'$ refer to the two primary weak
decay interactions of each diagram and can have the values $i \, (i')=n, \, p$;
where by $i \, (i')$ we mean
$\Lambda \, i \, (i') \rightarrow n \, i \, (i')$.
The remaining index, $j$, is the final state (\textit{i.e.} the emitted nucleons),
taking the values: $j$= $n$, $nn$, $np$, $nnn$, $nnp$, ...
For instance, in the example mentioned in the last section, we
have $i=i'=n$ and $j=np$. The first obvious constraint over
$j$ is the finite number of nucleons in any hypernuclei.
Also the charge is conserved and as a result there is always at least
one neutron in the final state.

The values for $\Gamma_{i, i' \rightarrow j}$ result from evaluating
any possible Goldstone diagram for the $\Lambda$-weak decay, where
the strong interaction is present.
We should be aware that $\Gamma_{i, i' \rightarrow j}$
does not represent only a decay width, but also some interference terms.
To understand the origin of these interference terms, the following
example can be instructive: we start with the
quantity $\Gamma_{n}$, which represents the decay width for the transition
amplitude $A_{1}: \, \Lambda n \rightarrow nn$ (for convenience,
we have numerated the transition amplitude). In fact, we evaluate the
square of this transition amplitude to obtain the final value $\Gamma_{n}$.
Once the FSI are incorporated, new transition amplitudes should be
included: let us consider the transition amplitude
$A_{2}: \, \Lambda p \rightarrow np \rightarrow nn$ \footnotemark{\footnotetext{
A more complete expression would be,
$A_{2}: \, \Lambda p \, (n_{b}) \rightarrow np \, (n_{b})  \rightarrow nn \, (p_{\, b})$
where the subindex $b$, indicates bound particles acting as spectators. In
this way, the charge conservation is more clearly seen.}},
where the strong interaction is responsible for the second reaction. This
transition amplitude is added to the first one and then the whole
expression is squared. From this simple model, four terms come out:
the squares of $A_{1}$ and $A_{2}$ and the two interference terms
between $A_{1}$ and $A_{2}$. The interference terms are non-zero
because the initial and the final states are the same.
The first two, can be interpreted as
decay widths, but not the interference terms which can be either
positive or negative. Besides this simple example,
by means of $\Gamma_{i, i' \rightarrow j}$ we
represent all terms but the squares of $\Lambda n \rightarrow nn$
and $\Lambda p \rightarrow np$.
A second important observation about the interference terms in
$\Gamma_{i, i' \rightarrow j}$, is that a particular case of these
terms are the ones between the $n$- and $p$-induced weak decays.
The interference terms between $A_{1}$ and $A_{2}$ are an
example of such contribution.
One of the simplest diagrams
for these interference terms, is the first ring diagram
in Fig.~1, taking one bubble as a $p p^{-1}$-pair
and the other one as a $n n^{-1}$-pair. Note also that
in this diagram the nuclear strong
interaction appears in first order and the sign of the contribution
depends on the sign of this interaction.
The interference terms in $\Gamma_{i, i' \rightarrow j}$
are pure quantum mechanical effects and the ones with $i \neq i'$,
are not contained in the semi-phenomenological INC calculation.

Now, we write down expressions for $N_{N}$,
\begin{eqnarray}
\label{factfnn}
N_{n} & = & 2 \bar{\Gamma}_{n} + \bar{\Gamma}_{p} + \,
\sum_{i, \, i'=n, \, p; \, j} N_{j \, (n)} \, \bar{\Gamma}_{i, i' \rightarrow j}, \\
\label{factfnp}
N_{p} & = & \bar{\Gamma}_{p} + \,
\sum_{i, \, i'=n, \, p; \, j} N_{j \, (p)} \, \bar{\Gamma}_{i, i' \rightarrow j},
\end{eqnarray}
and for $N_{NN}$,
\begin{eqnarray}
\label{factfnnn}
N_{nn} & = & \bar{\Gamma}_{n} + \,
\sum_{i, \, i'=n, \, p; \, j} N_{j \, (nn)} \, \bar{\Gamma}_{i, i' \rightarrow j}, \\
\label{factfnnp}
N_{np} & = & \bar{\Gamma}_{p} + \,
\sum_{i, \, i'=n, \, p; \, j} N_{j \, (np)} \, \bar{\Gamma}_{i, i' \rightarrow j}, \\
\label{factfnpp}
N_{pp} & = &
\sum_{i, \, i'=n, \, p; \, j} N_{j \, (pp)} \, \bar{\Gamma}_{i, i' \rightarrow j},
\end{eqnarray}
where the normalization is the same as in Eq.~(\ref{nn1}) ({\it i. e.}
$\bar{\Gamma}_{i, i' \rightarrow j} \equiv
\Gamma_{i, i' \rightarrow j}/(\Gamma_{n}+\Gamma_{p})$).
The factors $N_{j \, (N)}$ are the numbers of nucleons of the type $N$
in the state $j$. In the same way, $N_{j \, (NN)}$ are the numbers of
pairs of nucleons of the type $NN$ in the state $j$. For instance,
if $j=nn$, $N_{nn \, (n)}=2$, and $N_{nn \, (p)}=0$.

Expressions for $N^{\rm 1B \textit{i}}_{N}$ and $N^{\rm 1B\textit{i}}_{NN}$ are obtained
by comparison between Eqs.~(\ref{nn1}-\ref{npp1}) and
Eqs.~(\ref{factfnn}-\ref{factfnpp}). To this end, the
$\Gamma_{i, i' \rightarrow j}$-terms with $i \neq i'$ must be neglected.
We have then,
\begin{eqnarray}
\label{factn}
N^{\rm 1B\textit{i}}_{N} & = & (N^{\rm 1B\textit{i}}_{N})^{wd} \, + \, \sum_{j}
N_{j \, (N)} \, \frac{\Gamma_{i, i \rightarrow j}}{\Gamma_{i}} \\
\label{factnn}
N^{\rm 1B\textit{i}}_{NN} & = & (N^{\rm 1B\textit{i}}_{NN})^{wd} \, + \, \sum_{j}
N_{j \, (NN)} \, \frac{\Gamma_{i, i \rightarrow j}}{\Gamma_{i}},
\end{eqnarray}
where in order to reduce the expressions we have used the factors in Table~I.
As mentioned above, within the INC calculation these factors are
independent of the weak vertex. Our expressions fulfilled this
requirement only when the transition potential is reduced to
one term: in this case, the weak vertex constant is  simplified in
the ratio $\Gamma_{i, i \rightarrow j}/\Gamma_{i}$.
When the full transition potential is present, then
$\Gamma_{i, i \rightarrow j}/\Gamma_{i}$ has some
dependence on the weak vertex. This point is further analyzed
in Section~V.

Before we end this section, we would like to make a brief comment
on the calculations performed in finite hypernuclei. In this case,
the weak decay matrix element, required for $\Gamma_{n, \, (p)}$,
is evaluated using finite nucleus wave-functions (WF). These WF
can be the harmonic oscillator WF, but also some more elaborate WF:
the ones from a BCS-, Tamm-Dancoff-, RPA-calculations or
any other model. In any case, these WF represent bound states.
After the weak decay takes place, the unbound emitted particles
are modelled by nuclear-matter distorted plane-waves. Nuclear correlations
between the emitted particles are the FSI, which do not affect
the $\Gamma_{n, \, (p)}$-result. The FSI should not be confused
with nuclear correlations employed in the evaluation of the
finite nucleus WF. For instance, the finite nucleus WF can
be the RPA-ones and afterwards, the FSI can be implemented using
the RPA. In this case, there is no double-counting: first,
the RPA is performed within bound particles and then,
the nuclear correlations are taken into account between
the particles in the continuum.

\newpage
\section{THE RING APPROXIMATION}
\label{RING}

In this section we present expressions for the
quantities $\Gamma_{i, i' \rightarrow j}$, within the
ring approximation.
This is done in non-relativistic nuclear matter. Before we show these
expressions, it is convenient to outline the derivation of $\Gamma_n$ and $\Gamma_p$.
Even though these contributions have been already discussed in \cite{ba03}, we
repeat now the main points of that derivation,
because this simplifies our presentation. To be consistent with
the ring approximation, we neglect the Pauli exchange
term also in $\Gamma_{n \, (p)}$. In $\Gamma_{n \, (p)}$ only
the transition potential $V^{\Lambda N}$ is present, while in
the ring approximation, also the strong interaction $V^{N N}$,
is required. Both $V^{\Lambda N}$ and $V^{N N}$ are
parameterized as follows,
\begin{equation}
\label{intlnnn}
V^{\Lambda N (NN)} (q) = \sum_{\tau_{\Lambda (N)}=0,1}
 {\cal O}_{\tau_{\Lambda (N)}}
{\cal V}_{\tau_{\Lambda (N)}}^{\Lambda N (NN)} (q),
\end{equation}
where $q$ is the energy-momentum carried by the potential.
The isospin dependence is given by,
\begin{eqnarray}
\label{isos}
{\cal O}_{\tau_{\Lambda (N)}} =~~~~~
\left\{
\begin{array}{c}1,~~\mbox{for}~~\tau_{\Lambda (N)}=0\\
  \v{\tau}_1 \cdot \v{\tau}_2,~~\mbox{for}~~\tau_{\Lambda (N)}=1
\end{array}\right.
\end{eqnarray}
The values $\tau=0,1$ stand for the isoscalar
and isovector parts of the interaction, respectively.
The spin and momentum dependence of the transition potential is,
\begin{eqnarray}
\label{intln}
{\cal V}_{\tau_{\Lambda}}^{\Lambda N} (q) &
= &  (G_F m_{\pi}^2)  \; \{
S_{\tau_{\Lambda}}(q)  \; \v{\sigma}_1 \cdot \v{\hat{q}} +
S'_{\tau_{\Lambda}}(q)  \; \v{\sigma}_2 \cdot \v{\hat{q}} +
P_{L, \tau_{\Lambda}}(q)
 \v{\sigma}_1 \cdot \v{\hat{q}} \; \v{\sigma}_2 \cdot
\v{\hat{q}} + P_{C, \tau_{\Lambda}}(q)  +  \nonumber \\
& & +  P_{T, \tau_{\Lambda}}(q)  (\v{\sigma}_1 \times \v{\hat{q}})
\cdot  (\v{\sigma}_2 \times \v{\hat{q}}) + i S_{V, \tau_{\Lambda}}(q)
\v{(\sigma}_1 \times \v{\sigma}_2) \cdot
\v{\hat{q}} \},
\end{eqnarray}
where the quantities  $S_{\tau_{\Lambda}}(q)$, $S'_{\tau_{\Lambda}}(q)$,
$P_{L, \tau_{\Lambda}}(q)$, $P_{C,\tau_{\Lambda}}(q)$,
$P_{T, \tau_{\Lambda}}(q)$ and $S_{V, \tau_{\Lambda}}(q)$ contain
short range correlations (SRC) and  are given in Appendix B of \cite{ba03}.
They are built up from the full one-meson-exchange model,
which involves the  complete pseudoscalar and vector meson octets
($\pi,\eta,K,\rho,\omega,K^*$).
The $S$ ($P$)-terms are the parity violating (parity conserving)
terms of the transition potential.

The spin and momentum dependence of the nuclear residual interaction is drawn as,
\begin{equation}
\label{intnn2}
{\cal V}_{\tau_N}^{N N} (q)
= (\frac{f_{\pi}^2}{m_{\pi}^2})  \; \{
{\cal V}_{C, \,\tau_{N}}(q) +
{\cal V}_{L, \, \tau_{N}}(q)
\v{\sigma}_1 \cdot \v{\hat{q}} \; \v{\sigma}_2 \cdot
\v{\hat{q}} +
{\cal V}_{T, \, \tau_{N}}(q)
(\v{\sigma}_1 \times \v{\hat{q}})
\cdot  (\v{\sigma}_2 \times \v{\hat{q}}) \},
\end{equation}
where the functions ${\cal V}_{C, \,\tau_{N}}(q)$,
${\cal V}_{L, \, \tau_{N}}(q)$ and
${\cal V}_{T, \, \tau_{N}}(q)$ are adjusted to reproduce
any effective OME-nuclear residual interaction. In
addition, following the same procedure as for
${\cal V}_{\tau_{\Lambda}}^{\Lambda N} (q)$
(see~\cite{ba03}), SRC
can be incorporated into ${\cal V}_{\tau_N}^{N N} (q)$.

We go back to the evaluation of the $\Gamma$'s (both
$\Gamma_{n \, (p)}$ and $\Gamma_{i, i' \rightarrow j}$). It is more
suitable to work with a partial decay width
$\Gamma(k,k_{F_{n}}, k_{F_{p}})$, instead of $\Gamma$,
where, $k$ is the $\Lambda$ energy-momentum,
$k_{F_{n}}$ and $k_{F_{p}}$ are the Fermi momentum
for neutrons and protons, respectively.
To evaluate $\Gamma(k)$ for a particular nucleus one uses
either an effective Fermi momentum or the Local Density
Approximation (LDA) \cite{os85}. In this work the LDA is employed. In
this case, $k_{F_{n}}$ and $k_{F_{p}}$ become position-dependent and are defined as
$k_{F_{n \, (p)}}(r) = \hbar c (3 \pi^{2} \rho_{n \, (p)}(r)/2)^{1/3}$, where
$\rho_n(r)=\rho(r) N/(N+Z)$ and $\rho_p(r)=\rho(r) Z/(N+Z)$, with
$\rho(r)$, $N$ and $Z$ being, respectively, the density profile,
number of neutrons and number of protons of the nuclear core of the hypernucleus.
In the last case,
it is equivalent to write the function $\Gamma(k,k_{F_{n}}, k_{F_{p}})$
in terms of the densities as $\Gamma(k,\rho_n(r), \rho_p(r))$.
The LDA reads,
\begin{equation}
\label{decwpar3}
\Gamma(k) = \int d \v{r} \,
\Gamma(k,\rho_n(r), \rho_p(r)) \;  |\psi_{\Lambda}(\v{r})|^2 \,
\end{equation}
where for the $\Lambda$ wave function $\psi_{\Lambda}(\v{r})$, we
take the $1s_{1/2}$ wave function of a harmonic oscillator. This
decay width can be seen as the $k$-component of the $\Lambda$ decay width.
The total decay width is obtained by averaging over the $\Lambda$ momentum
distribution, $|\widetilde{\psi}_{\Lambda}(\v{k})|^2$, as follows,
\begin{equation}
\label{decwpar2}
\Gamma = \int d \v{k} \,
\Gamma(k) \;  |\widetilde{\psi}_{\Lambda}(\v{k})|^2 \,
\end{equation}
where $\widetilde{\psi}_{\Lambda}(\v{k})$ is the Fourier transform of
$\psi_{\Lambda}(\v{r})$ and
$k_{0}=E_{\Lambda}(\v{k})+V_{\Lambda}$, being $V_{\Lambda}$
the binding energy for the $\Lambda$.

\subsection{The evaluation of $\Gamma_{n}$ and $\Gamma_{p}$}
We give now expressions for the partial decay widths
$\Gamma_{n \, (p)}(k,k_{F_{n}}, k_{F_{p}})$.
We should evaluate the first diagram in Fig.~2 and 3, named as $\Gamma_{i}$.
\begin{figure}
\begin{center}
    \includegraphics[width = .5\textwidth]{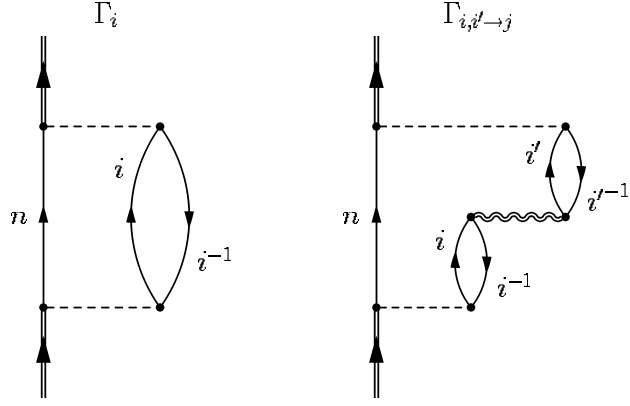}
\vskip 2mm
\caption{Goldstone diagrams for $\Gamma_{i}$ and
$\Gamma_{i, i' \rightarrow j}$, respectively. In both
diagrams, the vertex in the $\Lambda$-decay is
connected to a neutron. In the second diagram, the
double-wavy line is the dressed potential $\widetilde{V}^{N N}$,
which is described soon, in Eq.~(\ref{dyson}).}
\label{fig:fig2}
\end{center}
\end{figure}
\begin{figure}
\begin{center}
    \includegraphics[width = .5\textwidth]{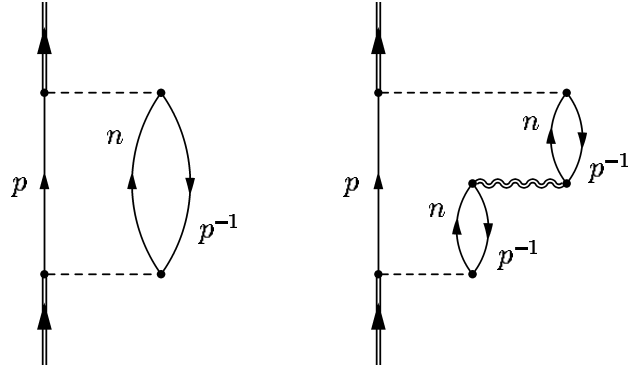}
\vskip 2mm
\caption{The same as in Fig.~2, but with a proton in the
$\Lambda$-decay vertex.}
\label{fig:fig2p}
\end{center}
\end{figure}
For convenience, in these figures we also present the ring contributions to
$\Gamma_{i, i' \rightarrow j}$.
We assign the names $p1$, $pi$ and $hi$ to the particle
between the $\Lambda$'s and the particle and hole in the bubble, respectively.
In order to perform the summation over spin and isospin quantum numbers, it is
more adequate to rewrite the transition rates in the form~\cite{ba03},
\begin{equation}
\label{gammtot}
\Gamma_{t_{hi}}(k,k_{F_{n}}, k_{F_{p}}) = \sum_{\tau,\tau'=0,1}
{\cal T}_{t_{hi}; \; \tau \tau'} \; \;
\widetilde{\Gamma}_{\tau \, \tau'}(k,k_{F_{n}}, k_{F_{p}})  \;
\end{equation}
where we distinguish
between the proton ($t_{hi}=1/2$) and the neutron induced ($t_{hi}=-1/2$) decay rates.
We have introduced the function,
\begin{equation}
\label{sumisosp}
{\cal T}_{t_{hi}; \, \tau \tau'}  =  \sum_{t_{p1}, t_{pi}}
\bra{t_\Lambda} {\cal O}_\tau \ket{t_{p1} t_{pi} t_{hi}}
\bra{t_{p1} t_{pi} t_{hi}} {\cal O}_{\tau'} \ket{t_\Lambda},
\end{equation}
to account for the isospin matrix elements.
The partial decay width as a function of the isospin of the transition
potential, is,
\begin{equation}
\label{gamdirl}
\widetilde{\Gamma}_{\tau \, \tau'}(k,k_{F_{n}}, k_{F_{p}})   =
(G_F m_{\pi}^2)^2 \frac{1}{(2 \pi)^2} \int d \v{p}_{1}
\theta(q_0) \theta(|\v{p}_{1}|-k_{Fp1})
\; {\cal S}_{\tau \tau'}(q)  \; (4 \, Im(\Pi^{0}t_{pi} \, t_{hi} (q))),
\end{equation}
where,
\begin{eqnarray}
\label{tdir}
{\cal S}_{\tau \tau'}(q) & = &  4 \; \{
S_{\tau}(q) S_{\tau'}(q) + S'_{\tau}(q) S'_{\tau'}(q)
 + P_{L, \tau}(q)  P_{L, \tau'}(q)  +  P_{C, \tau}(q) P_{C,
\tau'}(q) + \nonumber \\
 & & + 2  \, P_{T, \tau}(q)  P_{T, \tau'}(q) + 2 \, S_{V,
\tau}(q) S_{V, \tau'}(q) \}
\end{eqnarray}
and
\begin{equation}
\label{ring3}
\Pi^{0}t_{pi} \, t_{hi} (q)  =  \frac{-1~~~}{(2 \pi)^3} \; \int d \v{p}_{i}
\; \; \frac{\theta(|\v{p}_{i}|-k_{Fpi}) \theta(k_{Fhi}-
|\v{h}_{i}|)}
{q_0 - (E_{N}(\v{p}_{\, i}) - E_{N}( \v{h}_{i})) + i \eta},
\end{equation}
is the polarization propagator. Here $E_N$ is the nucleon total free energy.
The $(k,k_{F_{n}}, k_{F_{p}})$-dependence of the partial widths
$\widetilde{\Gamma}_{\tau \, \tau'}(k,k_{F_{n}}, k_{F_{p}}) $ is eliminated by means of the
Eqs.~(\ref{decwpar2}) and (\ref{decwpar3}).
By performing now the isospin summation, the final result from Eq. (\ref{gammtot}) is,
\begin{eqnarray}
\label{decnp}
\Gamma_n & = &  \widetilde{\Gamma}_{11}  +  \widetilde{\Gamma}_{00}  + \widetilde{\Gamma}_{01}
+\widetilde{\Gamma}_{10}
\nonumber \\
\Gamma_p & = & 5 \, \widetilde{\Gamma}_{11} +  \widetilde{\Gamma}_{00} -
(\widetilde{\Gamma}_{01}+\widetilde{\Gamma}_{10}),
\end{eqnarray}
where the presence of terms $\widetilde{\Gamma}_{\tau \, \tau'}$ with
$\tau \neq \tau'$ is a consequence of the restriction in the isospin sum.
For the non-mesonic decay width, $\Gamma_{NM}=\Gamma_n +\Gamma_p$, these terms
cancel out.

\subsection{The evaluation of $\Gamma_{i, i' \rightarrow j}$}
In the ring approximation, diagrams
with one particle-one hole bubbles
are summed up to infinite order. The lowest order contribution
to the ring series is of first order in the
strong interaction, $V^{NN}$ and it is shown as the first
ring diagram in Fig.~1.
To obtain all the other contributions,
the basic idea is to replace $V^{NN}$,
by a dressed interaction, $\widetilde{V}^{NN}$, obtained as a
result of the sum of the ring series. This process is sketched in
Fig.~4, which is a representation of the Dyson equation,
\begin{figure}
\begin{center}
    \includegraphics[width = .5\textwidth]{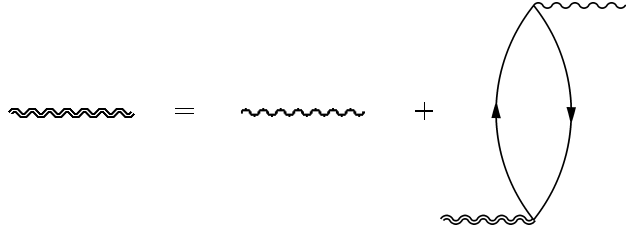}
\vskip 2mm
\caption{Diagrammatic representation of the Dyson equation. A
wavy line represents the nuclear interaction, $V^{NN}$, while
a double-wavy line is the dressed interaction $\widetilde{V}^{NN}$.}
\label{fig:fig3}
\end{center}
\end{figure}
\begin{equation}
\label{dyson}
\widetilde{{\cal V}}_{C,\, (L, \, T); \, \tau_{N}} =
{\cal V}_{C,\, (L, \, T); \, \tau_{N}}
 + \widetilde{{\cal V}}_{C,\, (L, \, T); \, \tau_{N}} \, C_{s, \, t }
 \; \Pi_{pi, \, hi}^{0} \;
 (f_{\pi}^2/m_{\pi}^2)  \,{\cal V}_{C,\, (L, \, T); \, \tau_{N}}
\end{equation}
where the constant $C_{s, \, t }$ comes from the sum over spin-isospin and
it is specified soon. We have shorten the notation for the polarization propagator.
The way in which the nuclear interaction is displayed in Eqs.~(\ref{intlnnn}) and (\ref{intnn2}),
has the advantage that there is no-interference between the central ($C$), spin-longitudinal ($L$)
and spin-transverse ($T$) terms, neither nor between the isoscalar ($\tau=0$) and
isovector ($\tau=1$) ones.
The solution for the Dyson equation is,
\begin{equation}
\label{dyson2}
\widetilde{{\cal V}}_{C,\, (L, \, T); \, \tau_{N}} =
\frac{{\cal V}_{C,\, (L, \, T); \, \tau_{N}}}{1 -
(f_{\pi}^2/m_{\pi}^2)  \,{\cal V}_{C,\, (L, \, T); \, \tau_{N}}
 \, C_{s, \, t } \, \Pi_{pi, \, hi}^{0}},
\end{equation}
The interaction $\widetilde{V}^{NN}$, is then,
\begin{equation}
\label{intlnnn2}
\widetilde{V}^{NN} (q) = \sum_{\tau_{N}=0,1}
 {\cal O}_{\tau_{N}}
\widetilde{{\cal V}}_{\tau_{N}}^{NN} (q),
\end{equation}
where ${\cal O}_{\tau_{N}}$ is defined in Eq.~(\ref{isos})
and
\begin{equation}
\label{intnn3}
\widetilde{{\cal V}}_{\tau_N}^{N N} (q)
= (\frac{f_{\pi}^2}{m_{\pi}^2})  \; \{
\widetilde{{\cal V}}_{C, \,\tau_{N}}(q) +
\widetilde{{\cal V}}_{L, \, \tau_{N}}(q)
\v{\sigma}_1 \cdot \v{\hat{q}} \; \v{\sigma}_2 \cdot
\v{\hat{q}} +
\widetilde{{\cal V}}_{T, \, \tau_{N}}(q)
(\v{\sigma}_1 \times \v{\hat{q}})
\cdot  (\v{\sigma}_2 \times \v{\hat{q}}) \}.
\end{equation}
The replacement of this interaction into the evaluation of
a decay width is not straightforward, because of the
restrictions in the isospin summation.
The isospin, is a crucial ingredient in the construction of
$\Gamma_{i, i' \rightarrow j}$.
Within the ring approximation, there are only two possible final states:
$nn$ or $np$. Therefore, we should give expressions for
eight quantities: $\Gamma_{n, n \rightarrow nn}$, $\Gamma_{n, p \rightarrow nn}$,
$\Gamma_{p, n \rightarrow nn}$, $\Gamma_{p, p \rightarrow nn}$ and the
remaining four expressions are obtained by replacing the final
state $nn$ by $np$ in the former four ones. In Figs.~5 and~6, we
show some of the lower order ring-diagrams which contribute
to $\Gamma_{i, i' \rightarrow nn}$ and $\Gamma_{i, i' \rightarrow np}$,
respectively.
\begin{figure}
\begin{center}
    \includegraphics[width = .7\textwidth]{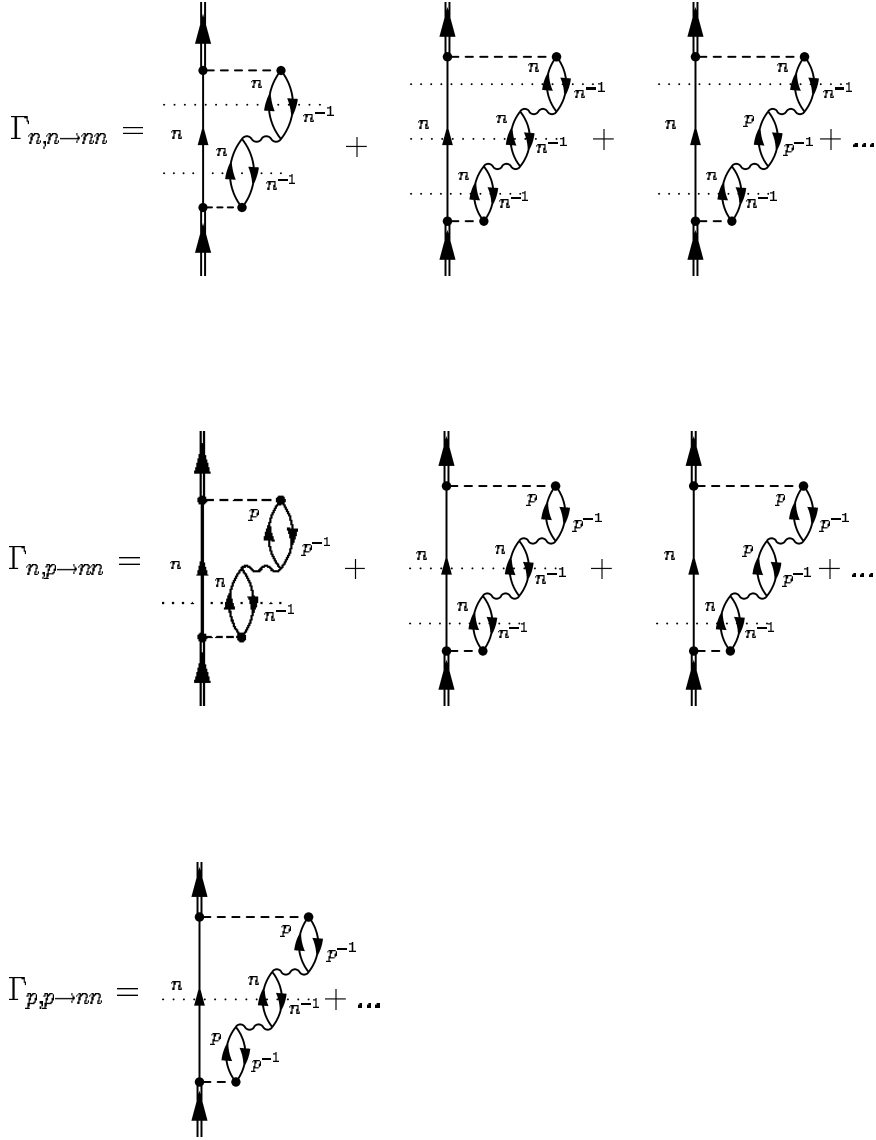}
\vskip 2mm
\caption{Goldstone diagrams for $\Gamma_{i, i' \rightarrow nn}$.
In this set of diagrams we present some representative examples of this quantity.
Here, a dotted line represents the state on the mass shell. A diagram with two or
more dotted lines stands for the sum of diagrams with one each.
The explicit inclusion of $\Gamma_{p, n \rightarrow nn}$ is omitted, as
it is analogous to $\Gamma_{n, p \rightarrow nn}$.}
\label{fig:fig5}
\end{center}
\end{figure}
\begin{figure}
\begin{center}
    \includegraphics[width = .7\textwidth]{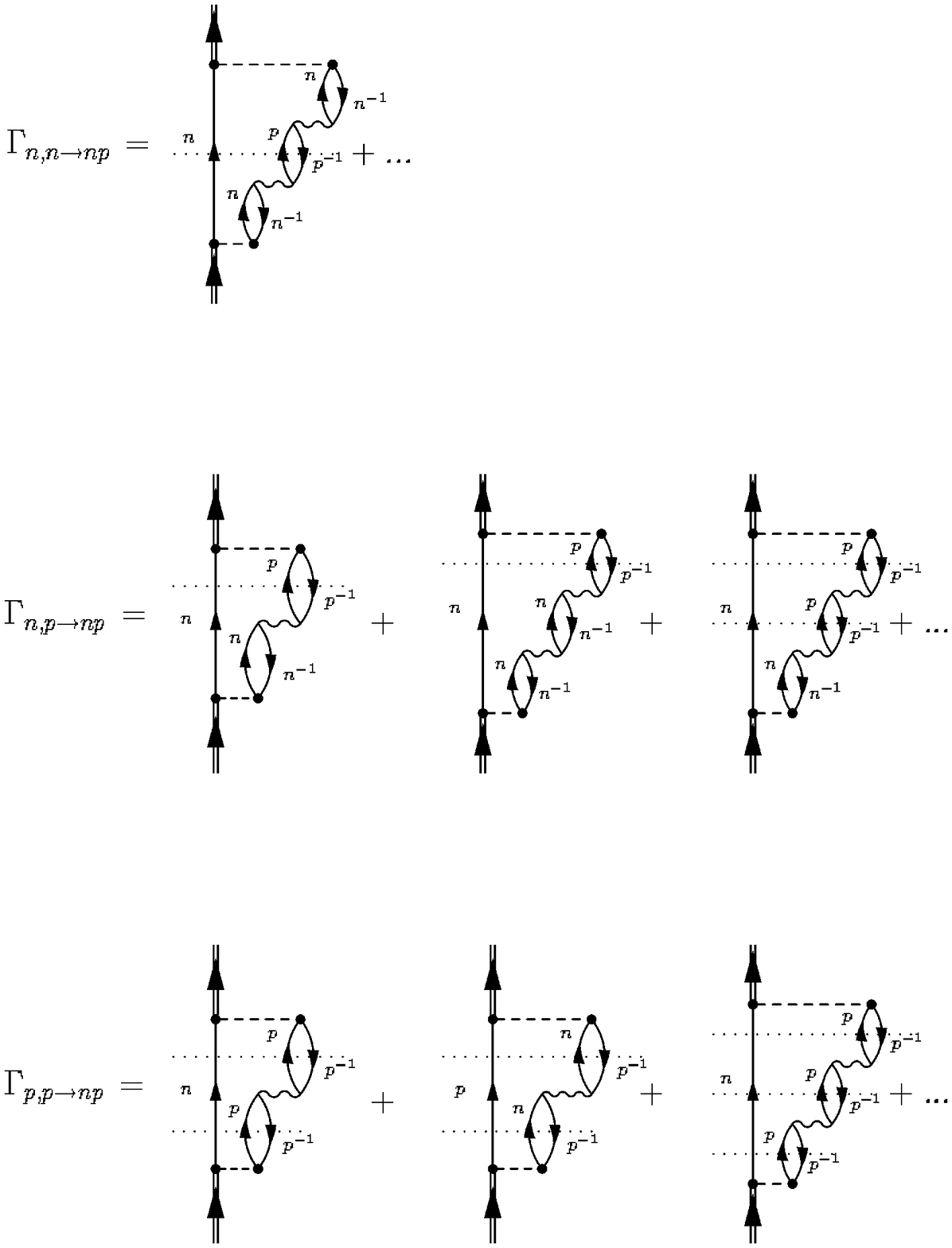}
\vskip 2mm
\caption{The same as in Fig.~5, but for $\Gamma_{i, i' \rightarrow np}$.}
\label{fig:fig6}
\end{center}
\end{figure}

To obtain general expressions for $\Gamma_{i, i' \rightarrow j}$, it is
more convenient to work with Figs.~2 and~3, where we show four
diagrams: the particle in the $\Lambda$-decay vertex is
a neutron in the first figure and a proton in the second case.
The particle-hole pairs $i i^{-1}$ and $i' i'^{-1}$
in Fig.~2, can take the values $nn^{-1}$ or $pp^{-1}$.
However, the isospin conservation allows only $n p^{-1}$-pairs
in the same position in Fig.~3. The same considerations hold
for the particle-hole pairs in $\widetilde{V}^{N N}$.
Consequently, by choosing adequately the particle-hole pairs
and the cut in the diagrams in Fig.~2, we have that the first one, contributes to
$\Gamma_{n}$ and $\Gamma_{p}$, while the second one does the same
to the eight $\Gamma_{i, i' \rightarrow j} \,$-functions. At variance,
the diagrams in Fig.~3 only play a role in $\Gamma_{p}$ and
in $\Gamma_{p, p \rightarrow np}$, for the first and second
diagrams, respectively.

Using as a guide the diagrams drawn in the second places
in Figs.~2 and 3, we build up
expressions for $\Gamma_{i, i' \rightarrow j}$.
We follow similar steps as in the derivation
of $\Gamma_{N}$. In this case, however, we begin
with a general expression, using
the standard Goldstone-rules for diagrams,
\begin{eqnarray}
\label{gamgold}
\Gamma_{t_{hi}, \, t_{hi'} \rightarrow j \;}(k,k_{F_{n}}, k_{F_{p}}) & = &
 -2 \, Im \, \int \frac{d^4 \, p_{1}}{(2 \pi)^4} \,
\int \frac{d^4 \, h_{i}}{(2 \pi)^4} \, \, \int \frac{d^4 \, h_{i'}}{(2 \pi)^4} \;
\frac{1}{4} \sum
G_{part}(p_{1}) G_{part}(p_{i})
\nonumber \\
&&
G_{part}(p_{i'}) G_{hole}(h_{i}) G_{hole}(h_{i'})
\nonumber \\
&& \times \bra{\gamma_{\Lambda}}
(V^{\Lambda N}(q))^{\dag}
\ket{\gamma_{p1} \gamma_{pi'} \gamma_{hi'}}
\bra{\gamma_{p1} \gamma_{pi'} \gamma_{hi'}}
\widetilde{V}^{N N}(q)
\ket{\gamma_{p1} \gamma_{pi} \gamma_{hi}}
\nonumber \\
&& \times \bra{\gamma_{p1} \gamma_{pi} \gamma_{hi}}
V^{\Lambda N}(q)
\ket{\gamma_{\Lambda}}.
\end{eqnarray}
For simplicity,
$\gamma_{l}$ represents the spin ($s$), isospin ($t$) and
energy-momentum of the particle $l$.
Although the meaning of the subindices $p i$ and $h i$
($pi'$ and $hi'$), refers to the particle-hole
pair $i i^{-1}$ ($i' i'^{-1}$) in Fig.~2, this
expression is also valid for the second diagram in Fig.~3.
Using the same notation as in Eq.~(\ref{gamdirl}), $p_{1}$
represents the particle between both $\Lambda's$. From the energy-momentum
conservation, we have, $q=k-p_{1}$, $p_{i}=q+h_{i}$ and $p_{i'}=q+h_{i'}$.
The summation runs over all spins,
while the isospin sum can not be specified until one sets the final state $j$.
The particle and the hole propagators are, respectively,
\begin{equation}
\label{nprop}
G_{part}(p)  =  \frac{\theta(|\mbox{\boldmath $p$}| - k_F)}
{p_0 - E_N(\mbox{\boldmath $p$})- V_N + i
\varepsilon}
\end{equation}
and
\begin{equation}
\label{nprop2}
G_{hole}(h)  = \frac{\theta(k_F - |\mbox{\boldmath $h$}| )}
{h_0 - E_N(\mbox{\boldmath $h$})- V_N - i
\varepsilon},
\end{equation}
where $V_N$ is the nucleon binding energy.
We re-write Eq.~(\ref{gamgold}) in terms of partial
decay widths as follows,
\begin{equation}
\label{gammtoti}
\Gamma_{i, i' \rightarrow j}(k,k_{F_{n}}, k_{F_{p}}) =
\sum_{\tau_{\Lambda},\tau_{N},\tau'_{\Lambda}=0,1}
{\cal T}_{t_{hi}, t_{hi'}; \; \tau_{\Lambda},\tau_{N},\tau'_{\Lambda}} \; \;
\widetilde{\Gamma}^{i, i' \rightarrow j}
_{\tau_{\Lambda},\tau_{N},\tau'_{\Lambda}}(k,k_{F_{n}}, k_{F_{p}})  \;
\end{equation}
where
\begin{equation}
\label{sumisospi}
{\cal T}_{t_{hi}, t_{hi'}; \; \tau_{\Lambda},\tau_{N},\tau'_{\Lambda}} \;
= \sum
\bra{t_\Lambda} {\cal O}_{\tau'_{\Lambda}} \ket{t_{p1} t_{pi'} t_{hi'}}
\bra{t_{p1} t_{pi'} t_{hi'}} {\cal O}_{\tau_{N}} \ket{t_{p1} t_{pi} t_{hi}}
\bra{t_{p1} t_{pi} t_{hi}} {\cal O}_{\tau_{\Lambda}} \ket{t_\Lambda}.
\end{equation}
Now the summation runs only over isospin, with the same
consideration as in Eq.~(\ref{gamgold}).
When performing the energy integrations in Eq.~(\ref{gamgold}), one
keeps only two-particles on the mass shell (more details are given soon).
The partial decay widths as function of the isospin of the transition
potential, are defined as,
\begin{equation}
\label{ring1}
\widetilde{\Gamma}_{\tau \, \tau_{N} \, \tau'}^{\; i, i' \rightarrow j, \, n}
(k,k_{F_{n}}, k_{F_{p}})   =
(G_F m_{\pi}^2)^2 (\frac{f_{\pi}^2}{m_{\pi}^2}) \; \frac{1}{(2 \pi)^2} \int d \v{p}_{1}
\theta(q_0) \theta(|\v{p}_{1}|-k_{F_{n}})
\; \widetilde{{\cal S}}^{i, i' \rightarrow j, \, n}_{\tau \, \tau_{N} \, \tau'}(q)
\end{equation}
for the second diagram in Fig.~2, where the super-index $n$ refers to
the particle in the $\Lambda$-vertex (which is explicitly indicated in
the step function). For the second diagram in Fig.~3, we have,
\begin{equation}
\label{ring1p}
\widetilde{\Gamma}_{\tau \, \tau_{N} \, \tau'}^{\; i, i'  \rightarrow j, \, p}
(k,k_{F_{n}}, k_{F_{p}})   =
(G_F m_{\pi}^2)^2 (\frac{f_{\pi}^2}{m_{\pi}^2}) \; \frac{1}{(2 \pi)^2} \int d \v{p}_{1}
\theta(q_0) \theta(|\v{p}_{1}|-k_{F_{p}})
\; \widetilde{{\cal S}}^{i, i'  \rightarrow j, \, p}_{\tau \, \tau_{N} \, \tau'}(q),
\end{equation}
where the meaning of the super-index $p$ is self-evident.
We define the ${\cal S}$-functions as (for details, see the Appendix),
\begin{eqnarray}
\label{tdir2}
\widetilde{{\cal S}}^{\; i, i'  \rightarrow j, \, n}_{\; \tau \, \tau_{N} \, \tau'}(q) & = &
-\{(S'_{\tau} S'_{\tau'} +
P_{C, \tau} P_{C, \tau'})
\widetilde{{\cal U}}^{\, i, i'  \rightarrow j, \, n}_{\; C; \, \tau_{N}}(q)+
(S_{\tau} S_{\tau'}
 + P_{L, \tau}  P_{L, \tau'})
\widetilde{{\cal U}}^{\, i, i'  \rightarrow j, \, n}_{ \; L; \, \tau_{N}}(q)
 + \nonumber \\
 & & + 2  \, ( S_{V, \tau} S_{V, \tau'} +
P_{T, \tau} P_{T, \tau'})
\widetilde{{\cal U}}^{\, i, i'  \rightarrow j, \, n}_{ \; T; \, \tau_{N}}(q)\}
\end{eqnarray}
and
\begin{eqnarray}
\label{tdir2p}
\widetilde{{\cal S}}^{\; i, i'  \rightarrow j, \, p}_{\; \tau \, \tau_{N} \, \tau'}(q) & = &
-\{(S'_{\tau} S'_{\tau'} +
P_{C, \tau} P_{C, \tau'})
\widetilde{{\cal U}}^{\, i, i'  \rightarrow j, \, p}_{\; C; \, \tau_{N}}(q)+
(S_{\tau} S_{\tau'}
 + P_{L, \tau}  P_{L, \tau'})
\widetilde{{\cal U}}^{\, i, i'  \rightarrow j, \, p}_{ \; L; \, \tau_{N}}(q)
 + \nonumber \\
 & & + 2  \, ( S_{V, \tau} S_{V, \tau'} +
P_{T, \tau} P_{T, \tau'})
\widetilde{{\cal U}}^{\, i, i'  \rightarrow j, \, p}_{ \; T; \, \tau_{N}}(q)\},
\end{eqnarray}
To construct the functions
$\widetilde{{\cal U}}^{\, i, i'  \rightarrow j, \, n}_{ \; C, \, (L, \,T); \, \tau_{N}}(q)$
and $\widetilde{{\cal U}}^{\, i, i'  \rightarrow j, \, p}_{
\; C, \, (L, \,T); \, \tau_{N}}(q)$,
we first interpret the particle-hole bubbles
and the residual interaction, $\widetilde{V}^{NN}$ (which
contains a sum of particle-hole bubbles), in terms of configurations
and the possible final states that they can generate.
The particle-hole bubble is represented by
the polarization propagator, $\Pi^{0}t_{p} \, t_{h} (q)$,
which has a real and an imaginary part. The physical state
in any diagram, is determined by the position where the cut of the
diagram is done. In practical terms and within the present
problem, the imaginary part of the polarization propagator produce
the final state, while the real part is related with configurations
off the mass shell.
By performing the sum over isospin, the polarization propagator is
replaced by the sum $(\Pi^{0}_{nn}+\Pi^{0}_{pp})$ for the
$\Gamma_{i, i' \rightarrow j}$-diagrams in Fig.~2, and
by $\Pi^{0}_{np}$ for the ones in Fig.~3.
Note that this is done not only for
the bubbles explicitly drawn in these figures,
but also for the ones inside $\widetilde{V}^{NN}$.

Having in mind these considerations, explicit expressions for the
interaction in Eq.~(\ref{dyson2}) are obtained as follows.
When $j=nn$ the diagrams in Fig.~3 have no contribution, while in Fig.~2,
the polarization propagator is replaced by,
$(Re \Pi_{nn}^{0} + Re \Pi_{pp}^{0} + i Im \Pi_{nn}^{0})$: we have only
retained the imaginary part from the $nn^{-1}$-bubble. This neutron,
together with the one produced in the $\Lambda$-decay vertex, give rise to
the final $nn$-state. The $pp^{-1}$-bubble has a contribution off the
mass shell and therefore we keep only the real part of it.
In an analogously way, if $j=np$, the polarization propagator is
$(Re \Pi_{nn}^{0} + Re \Pi_{pp}^{0} + i Im \Pi_{pp}^{0})$. The diagrams
in Fig.~3 contribute only to $j=np$ and in this case the polarization
propagator is replaced by,
$(Re \Pi_{np}^{0} + i Im \Pi_{np}^{0})$. From Eq.~(\ref{dyson2})
we have, then,
\begin{eqnarray}
\label{dyson2nn}
\widetilde{{\cal V}}_{nn, \, C,\, (L, \, T); \, \tau_{N}} & = &
\frac{{\cal V}_{C,\, (L, \, T); \, \tau_{N}}}{1 -
2 \, (f_{\pi}^2/m_{\pi}^2)  \, {\cal V}_{C,\, (L, \, T); \, \tau_{N}}
(Re \Pi_{nn}^{0} + Re \Pi_{pp}^{0} + i Im \Pi_{nn}^{0})},
\nonumber \\
\widetilde{{\cal V}}_{np, \, C,\, (L, \, T); \, \tau_{N}} & = &
\frac{{\cal V}_{C,\, (L, \, T); \, \tau_{N}}}{1 -
4 \, (f_{\pi}^2/m_{\pi}^2)  \, {\cal V}_{C,\, (L, \, T); \, \tau_{N}}
(Re \Pi_{np}^{0} + i Im \Pi_{np}^{0})} \;\;\;\;\;\;\;\; \mbox{and}
\nonumber \\
\widetilde{{\cal V}}_{pp, \, C,\, (L, \, T); \, \tau_{N}} & = &
\frac{{\cal V}_{C,\, (L, \, T); \, \tau_{N}}}{1 -
2 \, (f_{\pi}^2/m_{\pi}^2)  \, {\cal V}_{C,\, (L, \, T); \, \tau_{N}}
(Re \Pi_{nn}^{0} + Re \Pi_{pp}^{0} + i Im \Pi_{pp}^{0})}.
\end{eqnarray}

The next step is to study the configurations in the weak vertices.
In this case, we should look at the particle-hole bubbles explicitly
drawn in the \textit{r.h.s} of Fig.~2 and 3. For $\Gamma_{n, n \rightarrow nn}$,
these bubbles are replaced by $\Pi_{nn}^{0}$. For $\Gamma_{p, p \rightarrow nn}$,
the replacement is limited to $Re \Pi_{pp}^{0}$: as $j=nn$, all protons should
be off the mass shell. Finally, for $\Gamma_{n, p \rightarrow nn}$
(or $\Gamma_{p, n \rightarrow nn}$), one bubble is replaced by
$\Pi_{nn}^{0}$ and the other one by $Re \Pi_{pp}^{0}$. Using the same procedure
for $j=np$, we have,
\begin{eqnarray}
\label{ringii1}
\widetilde{{\cal U}}^{\; nn \rightarrow nn, \, n}_{\; C,\, (L, \, T); \, \tau_{N}}(q) & = &
(Re \{\Pi_{nn}^{0} \})^{2} \,
Im  \{\widetilde{{\cal V}}_{nn, \, C,\, (L, \, T); \, \tau_{N}} \} + \nonumber \\
&+& 2 \, Re \{\Pi_{nn}^{0} \} \, Im \{\Pi_{nn}^{0} \} \,
Re \{\widetilde{{\cal V}}_{nn, \, C,\, (L, \, T); \, \tau_{N}} \}
\nonumber \\
\widetilde{{\cal U}}^{\; nn \rightarrow np, \, n}_{\; C,\, (L, \, T); \, \tau_{N}}(q) & = &
(Re \{\Pi_{nn}^{0} \})^{2} \,
Im  \{\widetilde{{\cal V}}_{pp, \, C,\, (L, \, T); \, \tau_{N}} \}
\nonumber \\
\widetilde{{\cal U}}^{\; np \rightarrow nn, \, n}_{\; C,\, (L, \, T); \, \tau_{N}}(q) & = &
Re \{\Pi_{pp}^{0} \} \,
Im  \{\widetilde{{\cal V}}_{nn, \, C,\, (L, \, T); \, \tau_{N}}
\, \Pi_{nn}^{0}\}
\nonumber \\
\widetilde{{\cal U}}^{\; np \rightarrow np, \, n}_{\; C,\, (L, \, T); \, \tau_{N}}(q) & = &
Re \{\Pi_{nn}^{0} \} \,
Im  \{\widetilde{{\cal V}}_{pp, \, C,\, (L, \, T); \, \tau_{N}}
\, \Pi_{pp}^{0}\}
\nonumber \\
\widetilde{{\cal U}}^{\; pp \rightarrow nn, \, n}_{\; C,\, (L, \, T); \, \tau_{N}}(q) & = &
(Re \{\Pi_{pp}^{0} \})^{2} \,
Im  \{\widetilde{{\cal V}}_{nn, \, C,\, (L, \, T); \, \tau_{N}} \}
\nonumber \\
\widetilde{{\cal U}}^{\; pp \rightarrow np, \, n}_{\; C,\, (L, \, T); \, \tau_{N}}(q)
& = & (Re \{\Pi_{pp}^{0} \})^{2} \,
Im  \{\widetilde{{\cal V}}_{pp, \, C,\, (L, \, T); \, \tau_{N}} \} + \nonumber \\
&+& 2 \, Re \{\Pi_{pp}^{0} \} \, Im \{\Pi_{pp}^{0} \} \,
Re \{\widetilde{{\cal V}}_{pp, \, C,\, (L, \, T); \, \tau_{N}} \}
\nonumber \\
\widetilde{{\cal U}}^{\; pp \rightarrow np, \, p}_{\; C,\, (L, \, T); \, \tau_{N}}(q)
& = & (Re \{\Pi_{np}^{0} \})^{2} \,
Im  \{\widetilde{{\cal V}}_{np, \, C,\, (L, \, T); \, \tau_{N}} \} + \nonumber \\
&+& 2 \, Re \{\Pi_{np}^{0} \} \, Im \{\Pi_{np}^{0} \} \,
Re \{\widetilde{{\cal V}}_{np, \, C,\, (L, \, T); \, \tau_{N}} \}
\end{eqnarray}

Finally, the summation over isospin leads to,
\begin{eqnarray}
\label{decnp2}
\Gamma_{n, n \rightarrow nn} & = &
                \widetilde{\Gamma}^{n, n \rightarrow nn \, n}_{111} +
                \widetilde{\Gamma}^{n, n \rightarrow nn \, n}_{000} +
                \widetilde{\Gamma}^{n, n \rightarrow nn \, n}_{110} +
                \widetilde{\Gamma}^{n, n \rightarrow nn \, n}_{101} +
                \widetilde{\Gamma}^{n, n \rightarrow nn \, n}_{011} +
                \widetilde{\Gamma}^{n, n \rightarrow nn \, n}_{100} +
\nonumber \\
          &&
              + \widetilde{\Gamma}^{n, n \rightarrow nn \, n}_{010} +
                \widetilde{\Gamma}^{n, n \rightarrow nn \, n}_{001}
\nonumber \\
\Gamma_{n, p \rightarrow nn} & = &
                \widetilde{\Gamma}^{n, p \rightarrow nn \, n}_{111} +
                \widetilde{\Gamma}^{n, p \rightarrow nn \, n}_{000} +
                \widetilde{\Gamma}^{n, p \rightarrow nn \, n}_{110} -
                \widetilde{\Gamma}^{n, p \rightarrow nn \, n}_{101} -
                \widetilde{\Gamma}^{n, p \rightarrow nn \, n}_{011} -
                \widetilde{\Gamma}^{n, p \rightarrow nn \, n}_{100} -
\nonumber \\
          &&
              - \widetilde{\Gamma}^{n, p \rightarrow nn \, n}_{010} +
                \widetilde{\Gamma}^{n, p \rightarrow nn \, n}_{001}
\nonumber \\
\Gamma_{p, n \rightarrow nn} & = &
                \widetilde{\Gamma}^{p, n \rightarrow nn \, n}_{111}
              +  \widetilde{\Gamma}^{p, n \rightarrow nn \, n}_{000}
              -  \widetilde{\Gamma}^{p, n \rightarrow nn \, n}_{110}
              -  \widetilde{\Gamma}^{p, n \rightarrow nn \, n}_{101}
              +  \widetilde{\Gamma}^{p, n \rightarrow nn \, n}_{011}
              +  \widetilde{\Gamma}^{p, n \rightarrow nn \, n}_{100}-
\nonumber \\
          &&
              -  \widetilde{\Gamma}^{p, n \rightarrow nn \, n}_{010}
              -  \widetilde{\Gamma}^{p, n \rightarrow nn \, n}_{001}
\nonumber \\
\Gamma_{p, p \rightarrow nn} & = &
                \widetilde{\Gamma}^{p, p \rightarrow nn \, n}_{111} +
                \widetilde{\Gamma}^{p, p \rightarrow nn \, n}_{000} -
                \widetilde{\Gamma}^{p, p \rightarrow nn \, n}_{110} +
                \widetilde{\Gamma}^{p, p \rightarrow nn \, n}_{101} -
                \widetilde{\Gamma}^{p, p \rightarrow nn \, n}_{011} -
                \widetilde{\Gamma}^{p, p \rightarrow nn \, n}_{100} +
\nonumber \\
          &&
              + \widetilde{\Gamma}^{p, p \rightarrow nn \, n}_{010} -
                \widetilde{\Gamma}^{p, p \rightarrow nn \, n}_{001}
\nonumber \\
\Gamma_{n, n \rightarrow np} & = &
                \widetilde{\Gamma}^{n, n \rightarrow np \, n}_{111} +
                \widetilde{\Gamma}^{n, n \rightarrow np \, n}_{000} +
                \widetilde{\Gamma}^{n, n \rightarrow np \, n}_{110} +
                \widetilde{\Gamma}^{n, n \rightarrow np \, n}_{101} +
                \widetilde{\Gamma}^{n, n \rightarrow np \, n}_{011} +
                \widetilde{\Gamma}^{n, n \rightarrow np \, n}_{100} +
\nonumber \\
          &&
              + \widetilde{\Gamma}^{n, n \rightarrow np \, n}_{010} +
                \widetilde{\Gamma}^{n, n \rightarrow np \, n}_{001}
\nonumber \\
\Gamma_{n, p \rightarrow np} & = &
                \widetilde{\Gamma}^{n, p \rightarrow np \, n}_{111} +
                \widetilde{\Gamma}^{n, p \rightarrow np \, n}_{000} +
                \widetilde{\Gamma}^{n, p \rightarrow np \, n}_{110} -
                \widetilde{\Gamma}^{n, p \rightarrow np \, n}_{101} -
                \widetilde{\Gamma}^{n, p \rightarrow np \, n}_{011} -
                \widetilde{\Gamma}^{n, p \rightarrow np \, n}_{100} -
\nonumber \\
          &&
               -\widetilde{\Gamma}^{n, p \rightarrow np \, n}_{010} +
                \widetilde{\Gamma}^{n, p \rightarrow np \, n}_{001}
\nonumber \\
\Gamma_{p, n \rightarrow np} & = &
                 \widetilde{\Gamma}^{p, n \rightarrow np, \, n}_{111}
               + \widetilde{\Gamma}^{p, n \rightarrow np, \, n}_{000}
               - \widetilde{\Gamma}^{p, n \rightarrow np, \, n}_{110}
               - \widetilde{\Gamma}^{p, n \rightarrow np, \, n}_{101}
               + \widetilde{\Gamma}^{p, n \rightarrow np, \, n}_{011}
               + \widetilde{\Gamma}^{p, n \rightarrow np, \, n}_{100} -
\nonumber \\
          &&
               - \widetilde{\Gamma}^{p, n \rightarrow np, \, n}_{010}
               - \widetilde{\Gamma}^{p, n \rightarrow np, \, n}_{001}
\nonumber \\
\Gamma_{p, p \rightarrow np} & = &
                4 \, \widetilde{\Gamma}^{p, p \rightarrow np, \, p}_{111} +
                \widetilde{\Gamma}^{p, p \rightarrow np, \, n}_{111} +
                \widetilde{\Gamma}^{p, p \rightarrow np, \, n}_{000} -
                \widetilde{\Gamma}^{p, p \rightarrow np, \, n}_{110} +
                \widetilde{\Gamma}^{p, p \rightarrow np, \, n}_{101} -
                \widetilde{\Gamma}^{p, p \rightarrow np, \, n}_{011} -
\nonumber \\
          &&
               -\widetilde{\Gamma}^{p, p \rightarrow np, \, n}_{100} +
                \widetilde{\Gamma}^{p, p \rightarrow np, \, n}_{010} -
                \widetilde{\Gamma}^{p, p \rightarrow np, \, n}_{001}
\end{eqnarray}
The terms with 'mixed' isospin: we mean all terms but
$\widetilde{\Gamma}^{ii' \rightarrow j,  \, n (p)}_{111}$ and
$\widetilde{\Gamma}^{ii' \rightarrow j,  \, n (p)}_{000}$, are not
null because of the summation over the isospin is truncated. If
we sum up all terms with the same final state, these mixed
terms cancel out.

\subsection{Expressions for $N_{N}$ and $N_{NN}$}
From Eqs.~(\ref{factfnn}-\ref{factfnpp}), we finally have,
\begin{eqnarray}
\label{numnn}
N_{n}~ & = & 2 \, \bar{\Gamma}_{n} + \bar{\Gamma}_{p} +
2 \, (\bar{\Gamma}_{n, n \rightarrow nn} +
\bar{\Gamma}_{p, p \rightarrow nn} +
\bar{\Gamma}_{n, p \rightarrow nn} +
\bar{\Gamma}_{p, n \rightarrow nn}) +
\nonumber \\
&& + \bar{\Gamma}_{n, n \rightarrow np} +
\bar{\Gamma}_{p, p \rightarrow np} +
\bar{\Gamma}_{n, p \rightarrow np} +
\bar{\Gamma}_{p, n \rightarrow np}
\nonumber \\
N_{p}~ & = & \bar{\Gamma}_{p} +
\bar{\Gamma}_{n, n \rightarrow np} +
\bar{\Gamma}_{p, p \rightarrow np} +
\bar{\Gamma}_{n, p \rightarrow np} +
\bar{\Gamma}_{p, n \rightarrow np}
\nonumber \\
N_{nn} & = & \bar{\Gamma}_{n}  +
\bar{\Gamma}_{n, n \rightarrow nn} +
\bar{\Gamma}_{p, p \rightarrow nn} +
\bar{\Gamma}_{n, p \rightarrow nn} +
\bar{\Gamma}_{p, n \rightarrow nn}
\nonumber \\
N_{np} & = & \bar{\Gamma}_{p} +
\bar{\Gamma}_{n, n \rightarrow np} +
\bar{\Gamma}_{p, p \rightarrow np} +
\bar{\Gamma}_{n, p \rightarrow np} +
\bar{\Gamma}_{p, n \rightarrow np},
\end{eqnarray}
where we have employed again the normalization of Eq.~(\ref{nn1}).
For completeness, from Eqs.~(\ref{factn}-\ref{factnn}),
\begin{eqnarray}
\label{numdiag}
N^{\rm 1Bn}_{n} & = & 2 + \frac{2\, \Gamma_{n, n \rightarrow nn}
+ \Gamma_{n, n \rightarrow np}}{\Gamma_{n}}
\nonumber \\
N^{\rm 1Bn}_{p} & = & \frac{\Gamma_{n, n \rightarrow np}}{\Gamma_{n}}
\nonumber \\
N^{\rm 1Bp}_{n} & = & 1 + \frac{2\, \Gamma_{p, p \rightarrow nn}
+ \Gamma_{p, p \rightarrow np}}{\Gamma_{p}}
\nonumber \\
N^{\rm 1Bp}_{p} & = & 1 + \frac{\Gamma_{p, p \rightarrow np}}{\Gamma_{p}}
\nonumber \\
N^{\rm 1Bn}_{nn} & = & 1 + \frac{\Gamma_{n, n \rightarrow nn}}{\Gamma_{n}}
\nonumber \\
N^{\rm 1Bn}_{np} & = & \frac{\Gamma_{n, n \rightarrow np}}{\Gamma_{n}}
\nonumber \\
N^{\rm 1Bp}_{nn} & = & \frac{\Gamma_{p, p \rightarrow nn}}{\Gamma_{p}}
\nonumber \\
N^{\rm 1Bp}_{np} & = & 1 + \frac{\Gamma_{p, p \rightarrow np}}{\Gamma_{p}}
\end{eqnarray}

In the next section we give numerical results for these expressions.

\newpage

\section{RESULTS AND DISCUSSION}
\label{results}

In this section we present the  numerical results for the
ratios $N_{n}/N_{p}$ and $N_{nn}/N_{np}$, within the ring approximation.
We use nuclear matter in addition to the LDA which
allows us to discuss the $^{12}_{\Lambda}C$ hypernucleus.
As already mentioned, the transition potential is represented by the exchanges
of the $\pi$, $\eta$, $K$, $\rho$, $\omega$ and $K^*$-mesons.
We have employed several
nuclear residual interactions based on the OME.
In particular, we have used the Bonn potential~\cite{ma87}
in the framework of the parametrization presented in~\cite{br96},
which contains
the exchange of $\pi$, $\rho$, $\sigma$ and $\omega$ mesons,
while the $\eta$ and $\delta$-mesons are neglected.
In implementing the LDA,
the hyperon is assumed to be in the $1s_{1/2}$ orbit of  a
harmonic oscillator well with frequency
$\hbar \omega = 10.8$ MeV.
As already stated, we have employed different values for the proton and
neutron Fermi momenta, $k_{F_n}$ and $k_{F_p}$, respectively.

Before we present our results, we summarize four models for the
$V^{NN}$-strong potential:
\begin{itemize}
\item model 1: the Bonn potential~\cite{ma87} without SRC,

\item model 2: the Bonn potential~\cite{ma87} with  SRC,

\item model 3: a $(\pi+\rho)$-meson exchange potential with SRC.

\item model 4: a $(\pi+\rho)$-meson exchange potential without SRC, plus
the $g'$-Landau-Migdal parameter. In particular, we
have employed, $g'=0.5$ (in pionic units).
\end{itemize}

In Table~II we show values for $N_{n}/N_{p}$ and $N_{nn}/N_{np}$.
Our results underestimate the more recent data:
$N_{n}/N_{p} = 2.00 \pm 0.09 \pm 0.14$~\cite{ok04} and
$N_{nn}/N_{np}= 0.53 \pm 0.13$~\cite{kang06}.
These experimental points have an energy threshold of $T^{th}_{N}=60$~MeV
and 30~MeV, respectively. However,
a comparison with data is rather premature for reasons which are
discussed soon. From this table, we notice that the inclusion of
FSI improves the result. The SRC are important also for $V^{NN}$ and as usually stated,
this effect is well reproduced by the $g'$-Landau Migdal parameter. In this
table, the results for $\Gamma_{n}/\Gamma_{p}$ contain no Pauli-exchange terms, in order to
be consistent with the ring approximation. The inclusion of exchange terms
increases $\Gamma_{n}/\Gamma_{p}$ from 0.274 to 0.285~\cite{ba03}.
\begin{table}[h]
\begin{center}
\caption{
Results for the ratios $N_{n}/N_{p}$ and $N_{nn}/N_{np}$ for $^{12}_\Lambda C$.
In the first column we show the model for the nuclear interaction (see the text).
In the second and third columns we quote the values for the Fermi
momentum at $r=0$ (in units of MeV/c) for neutrons and protons, respectively.
In the columns $(N_{n}/N_{p})^{0}$ and $(N_{nn}/N_{np})^{0}$ we show the
results without FSI.
We have also included the values for $\Gamma_{n}/\Gamma_{p}$.}
\vspace{1cm}
\begin{tabular}{cccccccc}   \hline\hline
~~~$V^{NN}$~~~ & ~$k_{F_{n}}(r=0)$~ & ~$k_{F_{p}}(r=0)$~ &
~$\Gamma_{n}/\Gamma_{p}$~ & ~$(N_n/N_p)^{0}$~ &
~~$N_n/N_p$~~ & ~$(N_{nn}/N_{np})^{0}$~& ~~$N_{nn}/N_{np}$~~
\\  \hline
$model \; 1$ &  269.9  &  269.9  &  0.274 & 1.548 & 1.607 & 0.274 & 0.304  \\
$model \; 1$ &  261.5  &  277.8  &  0.229 & 1.458 & 1.514 & 0.229 & 0.257  \\
$model \; 2$ &  $idem$  &  $idem$  & $idem$  & $idem$ & 1.464 & $idem$ & 0.232  \\
$model \; 3$ &  $idem$  &  $idem$  & $idem$  & $idem$ & 1.462 & $idem$ & 0.231  \\
$model \; 4$ &  $idem$  &  $idem$  & $idem$
& $idem$ & 1.463 & $idem$ & 0.231  \\
\hline\hline \\
\end{tabular}
\end{center}
\end{table}

In Table~III we analyze the importance of the quantum-mechanical
interference terms terms between the
$\Lambda n \rightarrow nn$ and $\Lambda p \rightarrow np$ decay
channels. The interference terms are more relevant for the
$model \; 1$-interaction. The importance of the interference terms
when SRC are present, is reduced to approximately $1 \, \%$ for
$N_n/N_p$ and $2-3 \, \%$ for $N_{nn}/N_{np}$. When the SRC are
eliminated, these percentages increase up to $4 \, \%$ and
$12 \, \%$, respectively. The SRC produce a quasi-cancellation of
the interference terms. The trend, however, is that interference
terms are more significant for $N_{nn}/N_{np}$.
\begin{table}[h]
\begin{center}
\caption{
Results for the ratios $N_{n}/N_{p}$ and $N_{nn}/N_{np}$,
with and without the interference terms between the
$\Lambda n \rightarrow nn$ and $\Lambda p \rightarrow np$ amplitudes;
where the last ones are denoted with the $'diag'$ sub-index. The Fermi momentums at
$r=0$, are $k_{F_{n}}=261.5$ and $k_{F_{p}}=277.8$~MeV/c.}
\vspace{1cm}
\begin{tabular}{ccccc}   \hline\hline
~~~~~~~~~~~$V^{NN}$~~~~~~~~~~~ & ~~~~~$N_n/N_p$~~~~~ & ~~~$(N_n/N_p)_{diag}$~~~ &
~~~~~$N_{nn}/N_{np}$~~~~~ &
~~~$(N_{nn}/N_{np})_{diag}$~~~
\\  \hline
$model \; 1$ &  1.514  & 1.459   & 0.257  & 0.229   \\
$model \; 2$ &  1.464  & 1.452   & 0.232  & 0.226   \\
$model \; 3$ &  1.462  & 1.453   & 0.231  & 0.226   \\
\hline\hline \\
\end{tabular}
\end{center}
\end{table}

In Table~IV, we report values for
the $N^{\rm 1B\textit{i}}_{N, \, (NN)}$-factors
from Eqs.~(\ref{numdiag}). This is done for two $V^{\Lambda N}$-models:
the one with only one pion exchange
(OPE) and the complete OME.
Through this table, we
analyze the degree of dependence of these factors on the
weak vertex. Before this analysis,
we should note that the values for $N^{\rm 1Bn}_{p}$
and $N^{\rm 1Bp}_{p}$ are
identical to those for $N^{\rm 1Bn}_{np}$ and $N^{\rm 1Bp}_{np}$,
respectively. This is a consequence and a limitation of
the ring approximation, where the corresponding expressions
are the same (see Eqs.~(\ref{numdiag})). The value for
$\Gamma_{n}/\Gamma_{p}$ changes
from 0.167 (for OPE) to 0.229 (for OME), that is, a variation of $37 \, \%$,
while the same percentage is smaller for the more relevant
$N^{\rm 1B\textit{i}}_{N, \, (NN)}$-factors. However, it is
clear that within our model, these factors depend on the
weak vertex. To get a more clear understanding of the
effect of this dependence, in Table~V, we have employed
the experimental values for $N_{n}/N_{p}$ and
$N_{nn}/N_{np}$ together with the results in Table~IV
to get $\Gamma_{n}/\Gamma_{p}$ from the
Eqs.(\ref{gngpn}) and (\ref{gngpnn}).
We can see that the dependence on the weak vertex is negligible.
In order to avoid confusion, let us summarize the main
ideas. The INC calculation works in two sides: the theoretical
one, in which the $\Gamma_{n}/\Gamma_{p}$-result
depends on the weak vertex. And in second place, the search of a model
independent value for $\Gamma_{n}/\Gamma_{p}$, which is
called the experimental result for this ratio. This
value is built up from the experimental value for $N_{n}/N_{p}$ or
$N_{nn}/N_{np}$ and the results for
the $N^{\rm 1B\textit{i}}_{N, \, (NN)}$-factors, which
should be as model independent as possible. Our analytical
expressions for the $N^{\rm 1B\textit{i}}_{N, \, (NN)}$-factors
contradict the hypothesis of the independence of these factors
on the weak vertex, but our numerical results confirm the
possibility of a weak-vertex independent determination
of the ratio $\Gamma_{n}/\Gamma_{p}$, in agreement with
the INC calculation.
Also from Table~V, we can see that the values $(\Gamma_{n}/\Gamma_{p})_{Eq.~(\ref{gngpn})}$
and $(\Gamma_{n}/\Gamma_{p})_{Eq.~(\ref{gngpnn})}$ are consistent within each other.

\begin{table}[h]
\begin{center}
\caption{Results of the $N^{\rm 1B\textit{i}}_{N, \, (NN)}$-factors
within the ring approximation.
For the nuclear strong interaction we have taken the $model \, 1$ one,
where we have employed $k_{F_{n}}=261.5$ and $k_{F_{p}}=277.8$~MeV/c (at $r=0$).
In the line named as 'OPE' the $V^{\Lambda N}$-transition
potential is limited to a one pion-exchange.
As already stated, the OME contains
the  complete pseudoscalar and vector meson octets
($\pi,\eta,K,\rho,\omega,K^*$).}
\vspace{1cm}
\begin{tabular}{ccccc}   \hline\hline
~~~~~~~~~~~~~~$V^{\Lambda N}$~~~~~~~~~~~~~~ & ~~~~~$N^{\rm 1Bn}_{n}$~~~~~
& ~~~~~$N^{\rm 1Bn}_{p}$~~~~~
& ~~~~~$N^{\rm 1Bn}_{nn}$~~~~~
& ~~~~~$N^{\rm 1Bn}_{np}$~~~~~
\\  \hline
OPE &  2.315  &  0.060  & 1.127  &  0.060  \\
OME &  2.299  &  0.055  & 1.122  &  0.055  \\ \hline
$(\Delta N^{\rm 1B\textit{i}}_{N(N)}/N^{\rm 1B\textit{i}}_{N(N)}) \times 100$ &
 0.7   &  9.1  & 0.4  &  9.1  \\ \hline \hline
~~~~~~~~~~~$V^{\Lambda N}$~~~~~~~~~~~ & ~~~~~$N^{\rm 1Bp}_{n}$~~~~~
& ~~~~~$N^{\rm 1Bp}_{p}$~~~~~
& ~~~~~$N^{\rm 1Bp}_{nn}$~~~~~
& ~~~~~$N^{\rm 1Bp}_{np}$~~~~~
\\  \hline
OPE & 1.201  &  1.174  & 0.013  &  1.174  \\
OME & 1.156  &  1.141  & 0.011  &  1.141  \\ \hline
$(\Delta N^{\rm 1B\textit{i}}_{N(N)}/N^{\rm 1B\textit{i}}_{N(N)}) \times 100$ &
  3.9  &  2.9  & 18.2  & 2.9 \\
\hline\hline \\
\end{tabular}
\end{center}
\end{table}
\begin{table}[h]
\begin{center}
\caption{The ratio $\Gamma_{n}/\Gamma_{p}$ from Eqs.~(\ref{gngpn}-\ref{gngpnn}).
The values for the $N^{\rm 1B\textit{i}}_{N, \, (NN)}$-factors have been taken
from Table~IV and for $N_{n}/N_{p}$ and $N_{nn}/N_{np}$,
the data indicated in the text have been used.}
\vspace{1cm}
\begin{tabular}{ccc}   \hline\hline
~~~~~~~~~~~~~~$V^{\Lambda N}$~~~~~~~~~~~~~~ &
~~~~~~~~~~~~~~~~$(\Gamma_{n}/\Gamma_{p})_{Eq.~(\ref{gngpn})}$ ~~~~~
& ~~~~~~~~~~~~~~~~$(\Gamma_{n}/\Gamma_{p})_{Eq.~(\ref{gngpnn})}$ ~~~~~
\\  \hline
OPE & $0.52 \pm 0.08$ &  $0.56 \pm 0.14$ \\
OME & $0.51 \pm 0.07$ &  $0.54 \pm 0.14$ \\ \hline
$(\Delta (\Gamma_{n}/\Gamma_{p}) /(\Gamma_{n}/\Gamma_{p})) \times 100$
& $\sim 2$  & $\sim 3$  \\
\hline\hline \\
\end{tabular}
\end{center}
\end{table}

There are experimental information
not only for the ratios $N_{n}/N_{p}$ and $N_{nn}/N_{np}$
but also for the $N_{n}$ and $N_{p} \,$-spectra. In this contribution, we
have preferred not to present results for the spectra yet.
This is because, in developing a
microscopic scheme to interpret the data, we have presented and discussed a
formalism, where the numerical results are given
within the ring approximation. This implementation should be
seen as a first step towards more accurate results, which should
care about:
\begin{itemize}
\item the inclusion of the Pauli-exchange terms,

\item the addition of the two body-induced weak decays,

\item the FSI, which should not be restricted to the ring approximation and

\item the energy threshold used to obtain the data.
\end{itemize}
The Pauli-exchange terms. In \cite{ba03}, we have shown that
Pauli exchange terms are important.
In that work, a full antisymmetric calculation for the evaluation of
$\Gamma_{n}$ and $\Gamma_{p}$ has been done. As it is well known, the
ring approximation is the direct part of the RPA. In~\cite{ba96},
it has been established that the RPA-exchange terms are also
important. Even though the physical process studied in that work is different,
the kinematical conditions are similar: in both
cases the momentum carried by the nuclear interaction is of the
order of 400~MeV/c. From this fact, we can conclude that there is
a priori no reason to neglect the RPA-exchange terms. Nevertheless, the implementation
of the RPA in the present problem is quite an involved task. The
advantage of the ring series is that it can be summed up to infinite order.
Once exchange terms are incorporated, the RPA-series can not be summed
for a general finite range nuclear interaction and each term should
be computed individually. Even for the lowest order RPA-contribution,
there are three two-body operators: the nuclear strong interaction $V^{NN}$,
and the transition potential $V^{\Lambda N}$ (which appears twice). As for each
operator there is a direct and an exchange matrix element, one has to
evaluated eight terms. The RPA (or the lowest order contribution to it),
is beyond the scope of the present work, though a fair comparison with
data should certainly contain antisymmetrization.

Two body-induced decays. The non-mesonic weak decay of the $\Lambda$ is not only induced by
the process $\Lambda N \rightarrow NN$ but also by
the two-nucleon induced decay, $\Lambda NN \rightarrow NNN$.
If we call $\Gamma_{2}$ this decay width, then the total
non-mesonic decay width should be corrected as,
$\Gamma_{NM} = \Gamma_n + \Gamma_p + \Gamma_{2}$.
This last mechanism is originated from ground state correlations
and in~\cite{ba04}, we have shown that this decay
is $\sim 24 \, \%$ of the total non-mesonic decay width.
From these simple considerations it is clear that
the two body-induced decay plays a relevant role in the
evaluation of $N_{N}$ and $N_{NN}$, a point which has been already
discussed in~\cite{ba06}. The modifications in our expressions
to include this decay channel are simple. However,
the problem here is the huge number of possible diagrams which
should be considered.

The model for the FSI. The FSI admits two different approaches:
one is the semi-phenomenological INC calculation and the other one is a
microscopic treatment like ours. It is not possible to
establish a biunique relation between the processes in both models.
We focus on the microscopic point of view, and within this approach
the ring approximation constitutes a very particular set of
diagrams which contributes to the FSI.
There are also FSI beyond
the ring approximation which are important.
For instance, this issue is addressed in~\cite{al00},
where the bosonic loop expansion formalism is employed.
However, and as it is mentioned in that
very same work, within this scheme it is difficult to disentangle
two-nucleon from three-nucleon final states. Obviously,
the same consideration stated at the end of the last paragraph,
with respect to the number of possible diagrams, is held here.
In spite of these difficulties, we want to stress that the final
goal of our formalism (given by Eqs.~(\ref{factfnn})-(\ref{factfnpp})),
is the evaluation of FSI more complex than the ring approximation or the RPA.

The energy threshold. The $\Lambda$ within the
hypernucleus can decay either through the mesonic or through the
non-mesonic decay. In both cases, at least one nucleon is
produced. Obviously, the detectors can not distinguish
between nucleons coming from the mesonic and the non-mesonic decay.
Fortunately, the value for the energy of the nucleons from the mesonic decay is small.
Due to experimental conditions, up to now particles with energy smaller
than 30 MeV, can not be detected. This energy threshold is big
enough to eliminate all particles coming from the mesonic decay.
The numerical implementation of this
threshold is not difficult: in the expressions for the decay widths,
two step functions should be added to guarantee that the energy of the emitted
particles is greater than the threshold. In any case, this threshold
affects the neutron- and proton-spectra, more than the ratios
$N_{n}/N_{p}$ and $N_{nn}/N_{np}$. For this reason, together with the
will of making our presentation less complex, we have omitted its
inclusion.

From all these points, we wanted to emphasize that there are
several topics to overcome before we attempt a more realistic comparison
with data. The same consideration holds if we compare our results with
the ones from the INC calculation. Throughout the present contribution,
the INC calculation
has been taken as a reference guide. In spite of this, the INC calculation is
essentially different from our approach: in the INC calculation once the
$\Lambda$-decays, the trajectory of the emitted nucleons are followed
in their way out of the nucleus.
The free paths of these nucleons and their collisions cross sections,
are considered. From the four points mentioned above, the effect of the
second, third and fourth points are incorporated within the INC calculation. Due
to it semi-classical character, the first point together with some
interference terms can not be taken into account. At variance,
because of the quantum-mechanical character of our scheme, the concept
of trajectory is not applicable for us. In our case, by means of the
LDA, we allow the decay to take place in any point of the hypernucleus.
After the weak decay occurs, it follows the usual many-body problem; which
requires the election of certain sets of diagrams together with the election
of a nuclear residual interaction.
Here and as a first step, we have implemented the ring approximation.

\newpage

\section{CONCLUSIONS}
\label{CONCLU}
In this work we have addressed the problem of the
non-mesonic weak decay of $\Lambda$-
hypernuclei. As a result of the $\Lambda$-decay, nucleons
are emitted from the nucleus and
these nucleons can be measured. Moreover, nucleons from
the non-mesonic weak decay can
be disentangled from those originated in the mesonic decay.
We represent with $N_{N}$
and $N_{NN}$ the number of nucleons of kind $N$ and the number
of pairs of nucleons of
kind $NN$, respectively. We have developed for the first time
a microscopic model for $N_{N}$ and
$N_{NN}$, using non-relativistic nuclear matter. The interference
terms between the $n$- and $p$-induced weak decay widths are
automatically taken into account in our scheme.
Explicit expressions and numerical results are given
within the ring approximation together with the LDA.
Due to the limitations inherent
to the ring approximation, our numerical
results should be seen as preliminary. Our values
for the ratios $N_{n}/N_{p}$ and
$N_{nn}/N_{np}$ underestimate the data.
Our results do not
contain an energy threshold, but a threshold should
affect the $N_{n}$ and $N_{p}$-
spectrum, more than the mentioned ratios.
Within the same model, our value for
$\Gamma_{n}/\Gamma_{p}$ is in the range $0.2-0.3$.

It is our opinion that the so-called
$\Gamma_{n}/\Gamma_{p}$-puzzle has been misinterpreted.
In the literature, the theoretical predictions for
$\Gamma_{n}/\Gamma_{p}$ are smaller
than $0.5$. Old reported data for
this ratio are closer to one. More experiments
have been done and much theoretical effort
has been employed in the hope that the experiments produce
smaller values and/or theory
predicts bigger ones. The missing piece
of this puzzle is the link between theory and
experiment. The INC calculation address this point, but
it gives a different treatment to the weak
decay and to the subsequent nuclear many-body problem.
The INC calculation is not able to evaluate the interference terms
between the $n$- and $p$-induced weak decay widths.
In fact, these terms have never been evaluated before. Our
results show that these terms can be neglected when
SRC are present and within the ring approximation.

Besides these considerations, one question remains.
And it is if $\Gamma_{n}/\Gamma_{p}$
is an observable or not. Certainly, it is not
possible to perform a direct measurement of
the number of particles emitted in the
weak decay vertex: $\Gamma_{n}/\Gamma_{p}$ is obtained
indirectly. Now, if several models produce the
same $\Gamma_{n}/\Gamma_{p}$-result using the same experimental
information, then it would be fair to call this
number, the experimental result for $\Gamma_{n}/\Gamma_{p}$.
In the INC calculation, for instance,
the first step is the theoretical evaluation of the
$N^{\rm 1Bn \, (p)}_{n \, (p)}$-factors. Then, they are used
together with the data for $N_{n}/N_{p}$ (or $N_{nn}/N_{np}$),
in Eq.~(\ref{gngpn}) (or (\ref{gngpnn})) to obtain the
so-called experimental result for the ratio $\Gamma_{n}/\Gamma_{p}$ (which
is compared with the theoretical determinations of it).
Anyway, in the present contribution
we have tried to present an alternative point of view on this
subject. We calculate $N_{n}/N_{p}$ and $N_{nn}/N_{np}$ which are
straightaway compared with data.
Therefore, $\Gamma_{n}$ and $\Gamma_{p}$ are ingredients within our model. Regarding
the question about whether or not $\Gamma_{n}/\Gamma_{p}$ is observable,
is of secondary importance: once the data for $N_{n}/N_{p}$ and $N_{nn}/N_{np}$,
together with the corresponding spectra, are well understood, the disagreement
between the theoretical and the so-called experimental
value for $\Gamma_{n}/\Gamma_{p}$ would disappear automatically.

Then, as mentioned, we have presented
a unified model which evaluates
$\Gamma_{n}/\Gamma_{p}$, $N_{n}/N_{p}$ and
$N_{nn}/N_{np}$. This model is summarized in Eqs.~(\ref{factfnn})-(\ref{factfnpp}).
We have called attention on the fact that several
improvements over the ring approximation should be implemented before
we attempt a more realistic comparison with
data. From these points, the inclusion of the
Pauli exclusion principle should be the first
step. We believe that, in spite of the
difficulties inherent to the nuclear many-body problem,
we have shown that our proposal is feasible.
We also believe that the nuclear many-body problem after
the weak decay take place, is as important as the
weak decay itself.

\section*{Acknowledgments}
I would like to thank F. Krmpoti\'{c} for fruitful
discussions and for the critical reading of the
manuscript.
This work has been partially supported by the CONICET,
under contract PIP 6159.

\newpage

\section*{APPENDIX}
\label{APPEND}
In this Appendix we show some details about the derivation of the
$\widetilde{{\cal S}}^{\; i, i'  \rightarrow j, \, (n,p)}_{\; \tau \, \tau_{N} \, \tau'}(q)$-functions
(Eqs.~(\ref{tdir2})-(\ref{tdir2p})).
The starting point is Eq.~(\ref{gamgold}). By performing the energy integration, using the
factorization (\ref{gammtoti}) (which separates the isospin matrix elements),
the expressions for the transition potential
and the nuclear interaction given by Eqs.~(\ref{intln}) and (\ref{intnn3}), respectively
and after some algebra, we have,
\begin{eqnarray}
\label{ring1a1}
\widetilde{\Gamma}_{\tau \, \tau_{N} \, \tau'}^{\; i, i' \rightarrow j, \, n}
(k,k_{F_{n}}, k_{F_{p}})  & = &
\frac{-2~~}{(2 \pi)^9} \, \frac{1}{4}\, \sum_{all \, spin} \; Im \;
\int d \v{p}_{1} \int d \v{h}_{i} \int d \v{h}_{i'}
\theta(q_0) \theta(|\v{p}_{1}|-k_{F_{n}}) \nonumber \\
& & \times \frac{\theta(|\v{h}_{i} + \v{q}|-k_{Ft_{pi}})
\theta(k_{Ft_{hi}} - |\v{h}_{i}|)}{q_0 - (E_{N}(\v{h}_{i} + \v{q}) - E_{N}( \v{h}_{i})) + i \eta}
\nonumber \\
&& \times \frac{\theta(|\v{h}_{i'} + \v{q}|-k_{F t_{pi'}}) \theta(k_{F t_{hi'}} - |\v{h}_{i'}|)}
{q_0 - (E_{N}(\v{h}_{i'} + \v{q}) - E_{N}( \v{h}_{i'})) + i \eta}
\nonumber \\
&& \times \bra{s_{\Lambda}}
({\cal V}_{\tau}^{\Lambda N}(q))^{\dag}
\ket{s_{p1} s_{pi'} s_{hi'}}
\bra{s_{p1} s_{pi'} s_{hi'}}
\widetilde{{\cal V}}_{\tau_N}^{N N}(q)
\ket{s_{p1} s_{pi} s_{hi}}
\nonumber \\
&& \times \bra{s_{p1} s_{pi} s_{hi}}
{\cal V}_{\tau'}^{\Lambda N}(q)
\ket{s_{\Lambda}}.
\end{eqnarray}
and
\begin{eqnarray}
\label{ring1a2}
\widetilde{\Gamma}_{\tau \, \tau_{N} \, \tau'}^{\; i, i' \rightarrow j, \, p}
(k,k_{F_{n}}, k_{F_{p}})  & = &
\frac{-2~~}{(2 \pi)^9} \, \frac{1}{4}\, \sum_{all \, spin} \; Im \;
\int d \v{p}_{1} \int d \v{h}_{i} \int d \v{h}_{i'}
\theta(q_0) \theta(|\v{p}_{1}|-k_{F_{p}}) \nonumber \\
& & \times \frac{\theta(|\v{h}_{i} + \v{q}|-k_{F_{n}})
\theta(k_{F_{p}} - |\v{h}_{i}|)}{q_0 - (E_{N}(\v{h}_{i} + \v{q}) - E_{N}( \v{h}_{i})) + i \eta}
\nonumber \\
&& \times \frac{\theta(|\v{h}_{i'} + \v{q}|-k_{F_{n}}) \theta(k_{F_{p}} - |\v{h}_{i'}|)}
{q_0 - (E_{N}(\v{h}_{i'} + \v{q}) - E_{N}( \v{h}_{i'})) + i \eta}
\nonumber \\
&& \times \bra{s_{\Lambda}}
({\cal V}_{\tau}^{\Lambda N}(q))^{\dag}
\ket{s_{p1} s_{pi'} s_{hi'}}
\bra{s_{p1} s_{pi'} s_{hi'}}
\widetilde{{\cal V}}_{\tau_N}^{N N}(q)
\ket{s_{p1} s_{pi} s_{hi}}
\nonumber \\
&& \times \bra{s_{p1} s_{pi} s_{hi}}
{\cal V}_{\tau'}^{\Lambda N}(q)
\ket{s_{\Lambda}}.
\end{eqnarray}
where $k_{0}=E_{\Lambda}(\v{k})+V_{\Lambda}$ and $\v{q} = \v{p}_{1} - \v{k}$.
Now, the integrations over $\v{h}_{i}$ and $\v{h}_{i'}$ are replaced by the
function $\Pi^{0}t_{pi (i')} \, t_{hi(i')} (q)$ (Eq.~(\ref{ring3})).
Summing up in spin, we have,
\begin{eqnarray}
\label{ring1a3}
\widetilde{\Gamma}_{\tau \, \tau_{N} \, \tau'}^{\; i, i' \rightarrow j, \, n}
(k,k_{F_{n}}, k_{F_{p}})  & = & (G_F m_{\pi}^2)^2 (\frac{f_{\pi}^2}{m_{\pi}^2}) \,
\frac{-1~}{(2 \pi)^2} \; Im \;
\int d \v{p}_{1}
\theta(q_0) \theta(|\v{p}_{1}|-k_{F_{n}}) \Pi^{0}t_{pi} \, t_{hi} (q)
\nonumber \\
&& \times \Pi^{0}t_{pi'} \, t_{hi'} (q)
 \{(S'_{\tau} S'_{\tau'} +
P_{C, \tau} P_{C, \tau'})
\widetilde{{\cal V}}_{t_{p}t_{h}, \, C; \, \tau_{N}} +
(S_{\tau} S_{\tau'}
 + P_{L, \tau}  P_{L, \tau'})
\nonumber \\
&&
\times  \widetilde{{\cal V}}_{t_{p}t_{h}, \, L; \, \tau_{N}}
 +  2  \, ( S_{V, \tau} S_{V, \tau'} +
P_{T, \tau} P_{T, \tau'})
\widetilde{{\cal V}}_{t_{p}t_{h}, \, T; \, \tau_{N}} \}
\end{eqnarray}
and
\begin{eqnarray}
\label{ring1a4}
\widetilde{\Gamma}_{\tau \, \tau_{N} \, \tau'}^{\; i, i' \rightarrow j, \, p}
(k,k_{F_{n}}, k_{F_{p}})  & = & (G_F m_{\pi}^2)^2 (\frac{f_{\pi}^2}{m_{\pi}^2}) \,
\frac{-1~}{(2 \pi)^2} \; Im \;
\int d \v{p}_{1}
\theta(q_0) \theta(|\v{p}_{1}|-k_{F_{p}})
\; \Pi^{0}_{np} (q) \; \Pi^{0}_{np} (q)
\nonumber \\
&& \times \{(S'_{\tau} S'_{\tau'} +
P_{C, \tau} P_{C, \tau'})
\widetilde{{\cal V}}_{np, \, C; \, \tau_{N}} +
(S_{\tau} S_{\tau'}
 + P_{L, \tau}  P_{L, \tau'})
\widetilde{{\cal V}}_{np, \, L; \, \tau_{N}}
 + \nonumber \\
 & & + 2  \, ( S_{V, \tau} S_{V, \tau'} +
P_{T, \tau} P_{T, \tau'})
\widetilde{{\cal V}}_{np, \, T; \, \tau_{N}} \}
\end{eqnarray}
where the functions
$\widetilde{{\cal V}}_{t_{p}t_{h}, \, C (L,T); \, \tau_{N}} (q)$
are defined in Eq.~(\ref{dyson2nn}). The isospin is analyzed in detail
in Section~IV~B.

By comparison of Eqs.~(\ref{ring1a3}) and (\ref{ring1a4})
with Eqs.~(\ref{ring1}) and (\ref{ring1p}), respectively, it is straightforward
to obtain the expressions for
$\widetilde{{\cal S}}^{\; i, i'  \rightarrow j, \, n(p)}_{\; \tau \, \tau_{N} \, \tau'}(q)$
and $\widetilde{{\cal U}}^{\, i, i'  \rightarrow j, \, n(p)}_{\; C(L,T); \, \tau_{N}}(q)$.

\end{document}